\documentclass{article}
\usepackage[utf8]{inputenc}
\usepackage{authblk}
\usepackage{textcomp}
\usepackage{amsmath}
\usepackage{amssymb}
\usepackage{graphicx}
\usepackage{braket}
\usepackage{url} 
\usepackage{comment} 
\usepackage{subfigure}
\usepackage[makeroom]{cancel}
\usepackage{tikz}
\usetikzlibrary{calc}
\usetikzlibrary{arrows}
\usepackage{tikz-3dplot}

\usepackage{url}
\makeatletter
\def\url@leostyle{%
    \def\UrlFont{\sf}}{\def\UrlFont{\small\ttfamily}}
\makeatother
\usepackage[backend=biber,style=phys,]{biblatex}
  \bibliography{refs.bib}

\begin{document}

\title{\huge The Generative Programs Framework}

\author[1,2,a,*]{Mordecai Waegell}
\author[1,2,3,b]{Kelvin J. McQueen}
\author[1,2,3,4,c]{\newline Emily C. Adlam} 

\affil[1]{\small{Institute for Quantum Studies, Chapman University, Orange, CA 92866, USA}}

\affil[2]{Schmid College of Science and Technology, Chapman University, Orange, CA 92866, USA}

\affil[3]{Philosophy Department, Chapman University, Orange, CA 92866, USA}

\affil[4]{The Rotman Institute of Philosophy, University of Western Ontario,\newline 1151 Richmond Street, London, Ontario, Canada, N6A 5B7}

\affil[a]{ORCID: 0000-0002-1292-6041}

\affil[b]{ORCID: 0000-0003-3157-5168}

\affil[c]{ORCID: 0000-0002-5998-7685}

\affil[*]{Corresponding Author, email: waegell@chapman.edu}

\date{\today}

\maketitle

\textbf{Abstract:}
Recently there has been significant interest in using causal modelling techniques to understand the structure of physical theories. 
However, the notion of `causation' is limiting - insisting that a physical theory must involve causal structure already places significant constraints on the form that theory may take. Thus in this paper, we aim to set out a more general structural framework. We argue that any quantitative physical theory can be represented in the form of a  \textit{generative program}, i.e. a  list of instructions showing how to generate the empirical data; the information-processing structure associated with this program can be represented by a directed acyclic graph (DAG). We suggest that these graphs can be interpreted as encoding  relations of `ontological priority,' and that ontological priority is a suitable generalisation of causation which applies even to theories that don't have a natural causal structure.  We discuss some applications of our framework to philosophical questions about realism, operationalism, free will, locality and fine-tuning.

\section{Introduction}

Recently the quantum foundations community has been exploring the possibility of using causal models encoded in directed acyclic graphs (DAG) to study the structure of quantum mechanics \cite{SpekkensWood,schmid2020unscrambling}. This approach is based on the classical causal modelling framework advocated in particular by Pearl \cite{pearl2009causality}, Spirtes, Glymour, and Scheines \cite{spirtes2000causation}, and Woodward \cite{woodward2005making}, which  has subsequently been generalized to a quantum causal modelling framework making use of quantum operations and quantum channels \cite{Chaves,Miklin_2017,Pienaar_2017,https://doi.org/10.48550/arxiv.1906.10726}. Its proponents sometimes argued that realism, by definition, must take the form of assertions about causal structure \cite{schmid2020unscrambling}.

It is not hard to understand the appeal of the causal modelling approach: it is precise and quantitative and offers a  way of thinking about quantum mechanics which is clearly `realist' in the sense that it aims to explain rather than merely describe physical phenomena. However, causal modelling was developed for the study of macroscopic causation, not the microscopic world, and there are serious questions about whether it can reasonably be extrapolated to the microscopic world. In particular, fundamental physics shows almost no sign of the asymmetry that is usually taken to characterise the causal relation, and a number of credible accounts of causation suggest that it arises only at the macroscopic level, as a result of  the thermodynamic gradient ( \cite{Lewis1979-LEWCDA,10.2307/20012433}), interventions ( \cite{pearl2009causality,woodward2005making}), the perspectives of agents ( \cite{Price2005-PRICP}), or some such feature of reality which plays no role in fundamental physics.  So it could be the case that causal DAGs are simply the wrong tool to use to describe fundamental physics.   Woodward and Pearl themselves have both sounded a note of caution around the use of causal modelling in quantum mechanics - Pearl considers that the approach should only be applied to deterministic phenomena, and meanwhile Hausman and Woodward write, `\emph{If the measurement results are not distinct events or do not result from distinct mechanisms, then it must be mistaken to think of the EPR experiment as having a common cause structure or as having a structure in which the measurement results are related as cause and effect}' \cite{doi:10.1093/bjps/50.4.521}. Thus we consider that there is a clear need for a more general framework - one which doesn't presuppose an underlying causal structure, but which still allows us to offer precise, quantitative explanations of physical phenomena.   Such a framework should be able to accommodate both causal and non-causal approaches, in order that we can unify all such approaches in a common framework and provide definitions and arguments which apply equally well to all of them.

In this article we set out to provide such a framework, based on the notion of a `generative program' which can be represented as a DAG. Any theory which describes empirical regularities can be written in the form of a generative program, even if it is a purely operational theory. However, like the DAGs used in causal models, these DAGs can also be interpreted as putative descriptions of ontological structures: specifically, we suggest that they can be understood as encoding the relation of `ontological priority,' which has similarities with causation but is significantly more general. Thus we suggest  that causal modelling and causal language in the foundations of physics should generically be replaced by accounts in terms of `ontological priority,' with causal terminology being used judiciously only in instances where a causal account is clearly appropriate. 

After introducing the framework, we demonstrate several useful applications of it. First, we show how   generative programs are interpreted differently by operationalists and scientific realists, thus providing a clear characterisation of scientific realism which does not lean on assumptions about ontology or causal structure.  Second, we discuss how notions like `locality' and `contextuality' are represented in the generative program framework, and thus argue that these terms make sense only within a realist approach to physics - theories construed purely operationally cannot be local or non-local, and they cannot be contextual or non-contextual. Third, we discuss how the generative program framework can help clarify long-standing questions about free will and how it relates to physics.

\section{Generative Programs \label{programs}}

Our aim is that the generative program framework should be capable of accommodating all physical theories and interpretations of physical theories. We will understand a physical theory to be any formalism which generates some empirical data, perhaps conditional on a set of inputs; our framework thus excludes putative `physical theories' which are not capable of specifying any empirical data, but we would contend that if these theories do not eventually become capable of generating some empirical data, they should not be considered physical theories. It may also exclude some putative `physical theories' which are qualitative rather than quantitative, since such theories may be incapable of predicting concrete outcomes for specific scenarios, but since modern physics is largely expressed in quantitative terms this will not be a problem for most physical theories relevant to current science. Finally, we note that one may also think of a physical theory as simply `summarising' rather than `generating' the empirical data, but we will consider summarising data to be simply a special case of generating empirical data, since a summary can be thought of as a single massively conjunctive instruction to generate a set of known empirical data.

Evidently any such physical theory must provide (at least one) set of instructions explaining how the empirical data is to be generated, possibly conditional on a set of inputs. We will refer collectively to the instruction set and inputs as a `generative program.' In many cases a theory may provide several different ways of generating the same empirical data, so the theory can be associated with more than one generative program - for example, classical mechanics can be formulated in either a Hamiltonian or Lagrangian form, which would correspond to different generative programs. We put no  restrictions on the form that a generative program could take - in particular, we do not insist that it involves a set of dynamical laws which take in an initial state and perform time evolution to produce later states, and therefore this framework can be used even for `theories' which look very different from paradigmatic examples of physical theories based on forwards-in-time dynamical evolution. Even a physical theory which simply lists the data can still be understood as a generative program, albeit not a very interesting one because such a program does nothing to explain the regularities in the data.  Any set of logical instructions that generates the correct empirical data is a viable program within the framework, and because all such programs generate the same empirical data, they are empirically indistinguishable by definition.

A generative program will generally take the form of an ordered series of steps which are performed in the process of generating empirical data,  so it can be represented as a directed graph which shows the flow of information from inputs to outputs.  These graphs will typically be acyclic, because inconsistencies can arise if we are required to use the output of a later step in order to carry out an earlier step - in particular, we can easily run into logical contradictions similar to the grandfather paradox. In some cases there may be mathematical approaches which can get around these problems (for example, in the study of closed causal loops \cite{PhysRevD.44.3197,PhysRevD.46.4470} and process matrices \cite{Oreshkov2,Oreshkov,Ara_jo_2017,rubinorozema,Goswami_2018} it is sometimes possible to find a unique consistent solution for events happening in such a loop), and we hope to address these possibilities in later work, but for simplicity in this paper we will assume that we are dealing only with programs which can be represented by directed acyclic graphs. 

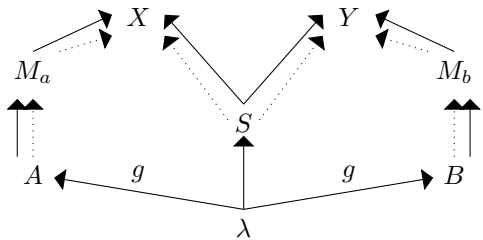
\begin{figure}[!ht]
		\centering
		\begin{tikzpicture}[scale=0.7]

		\coordinate (s) at (0,0);
			\coordinate (sm) at (0.3,-0.3);
				\coordinate (sp) at (-0.3,-0.3);
			\coordinate (smin) at (0,-0.6);
			\node[below, black] at (s) { $S$}; 
		\coordinate (y) at (2,2);
			\node[below, black] at (y) { $Y$}; 
				\coordinate (yp) at (1.5,1.7); 
					\coordinate (yp2) at (1.5,1.3);
						\coordinate (yp3) at (2.5,1.3); 
			\coordinate (ym) at (2.5,1.7); 
		\coordinate (mb) at (4,1); 
			\coordinate (mbm2) at (4,0.1);
				\coordinate (mb3) at (3.5,1); 
			\node[below, black] at (mb) { $M_b$}; 
		\coordinate (b) at (4,-1); 
			\node[below, black] at (b) { $B$}; 
		\coordinate (l) at (0,-2); 
	\node[below, black] at (l) { $\lambda$}; 
				\coordinate (ma) at (-4,1); 
						\coordinate (mam) at (-3.6,-1.4);
		\coordinate (mbm) at (3.6,-1.4);
					\coordinate (mam2) at (-4,0.1);
						\coordinate (ma3) at (-3.5,1); 
		\node[below, black] at (ma) { $M_a$}; 
			\coordinate (x) at (-2,2); 
			\coordinate (xp) at (-1.5,1.7); 
			\coordinate (xp2) at (-1.5,1.3); 
				\coordinate (xm) at (-2.5,1.7); 
				\coordinate (xp3) at (-2.5,1.3); 
		\node[below, black] at (x) { $X$}; 
		\coordinate (a) at (-4,-1); 
		\node[below, black] at (a) { $A$};

					\draw[black, arrows={-triangle 90}] (l) -- (mam);
		\draw[black, arrows={-triangle 90}] (l) -> (smin);
		\draw[black, arrows={-triangle 90}] (l) -- (mbm);
		\draw[black, arrows={-triangle 90}] (s) -- (yp);
		\draw[black, arrows={-triangle 90}] (s) -- (xp);
		\draw[black, arrows={-triangle 90}] (ma) -- (xm);
		\draw[black, arrows={-triangle 90}] (mb) -- (ym);
		
		\draw[black, dotted, arrows={-triangle 90}] (sm) -- (yp2);
		\draw[black, dotted, arrows={-triangle 90}] (sp) -- (xp2);

			\coordinate (a) at (-4,-1); 
					\coordinate (mam2) at (-4,0.1);
		\draw[black, dotted, arrows={-triangle 90}] (a) -- (mam2);
			\coordinate (mam22) at (-4.3,0.1);
				\coordinate (a22) at (-4.3,-1);
			\draw[black,  arrows={-triangle 90}] (a22) -- (mam22);

			\coordinate (ab) at (4,-1); 
					\coordinate (mam2b) at (4,0.1);
		\draw[black, dotted, arrows={-triangle 90}] (ab) -- (mam2b);
			\coordinate (mam22b) at (4.3,0.1);
				\coordinate (a22b) at (4.3,-1);
			\draw[black,  arrows={-triangle 90}] (a22b) -- (mam22b);

			\draw[black, dotted, arrows={-triangle 90}] (ma3) -- (xp3);
				\draw[black, dotted, arrows={-triangle 90}] (mb3) -- (yp3);

					\node[below, black] at (2,-1) { $g$}; 
						\node[below, black] at (-2,-1) { $g$}; 
		
		\end{tikzpicture}	
		\caption{ Both the ontic state of the entangled particles and the choices of measurement directions are determined by some hidden variable $\lambda$  (so statistical independence is violated).}
		\label{gt1}
	\end{figure}

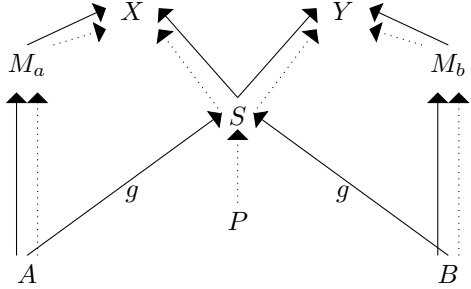
\begin{figure}[!ht]
	\centering
	\begin{tikzpicture}[scale=0.7]
	 
	\coordinate (s) at (0,0);
	\coordinate (sm) at (0.3,-0.3);
	\coordinate (sp) at (-0.3,-0.3);
	\coordinate (smin) at (0,-0.6);
	\node[below, black] at (s) { $S$}; 
	\coordinate (y) at (2,2);
	\node[below, black] at (y) { $Y$}; 
	\coordinate (yp) at (1.5,1.7); 
	\coordinate (yp2) at (1.5,1.3);
	\coordinate (yp3) at (2.5,1.3); 
	\coordinate (ym) at (2.5,1.7); 
	\coordinate (mb) at (4,1);
	\coordinate (mbm) at (4,0.2);
	\coordinate (mbm2) at (4,0.1);
	\coordinate (mbm23) at (3.8,0.1);
	\coordinate (mbm24) at (4.2,0.1);
	\coordinate (mb3) at (3.5,1); 
	\node[below, black] at (mb) { $M_b$}; 
	\coordinate (b) at (4,-3); 
	\coordinate (b4) at (4.2,-3); 
	\coordinate (b3) at (3.8,-3); 
	\node[below, black] at (b) { $B$}; 
	\coordinate (l) at (0,-2); 
	\node[below, black] at (l) { $P$}; 
	\coordinate (ma) at (-4,1); 
	\coordinate (mam) at (-4,0.2);
	\coordinate (mam2) at (-4,0.1);
		\coordinate (mam23) at (-4.2,0.1);
			\coordinate (mam24) at (-3.8,0.1);
	\coordinate (ma3) at (-3.5,1); 
	\node[below, black] at (ma) { $M_a$}; 
	\coordinate (x) at (-2,2); 
	\coordinate (xp) at (-1.5,1.7); 
	\coordinate (xp2) at (-1.5,1.3); 
	\coordinate (xm) at (-2.5,1.7); 
	\coordinate (xp3) at (-2.5,1.3); 
	\node[below, black] at (x) { $X$}; 
	\coordinate (a) at (-4,-3);
	\coordinate (a3) at (-4.2,-3);
		\coordinate (a4) at (-3.8,-3);
	\node[below, black] at (a) { $A$};

	\draw[black,dotted, arrows={-triangle 90}] (l) -> (smin);
	
		\draw[black, arrows={-triangle 90}] (a) -- (sp);
	\draw[black, arrows={-triangle 90}] (b) -- (sm);
	\draw[black, arrows={-triangle 90}] (s) -- (yp);
	\draw[black, arrows={-triangle 90}] (s) -- (xp);
	\draw[black, arrows={-triangle 90}] (ma) -- (xm);
	\draw[black, arrows={-triangle 90}] (mb) -- (ym);
	
	\draw[black, dotted, arrows={-triangle 90}] (sm) -- (yp2);
	\draw[black, dotted, arrows={-triangle 90}] (sp) -- (xp2);

	\draw[black, dotted, arrows={-triangle 90}] (a4) -- (mam24);
	
	\draw[black, dotted, arrows={-triangle 90}] (b4) -- (mbm24);

	\draw[black,   arrows={-triangle 90}] (a3) -- (mam23);
	
	\draw[black,   arrows={-triangle 90}] (b3) -- (mbm23);

	\draw[black, dotted, arrows={-triangle 90}] (ma3) -- (xp3);
	\draw[black, dotted, arrows={-triangle 90}] (mb3) -- (yp3);

	\node[below, black] at (2,-1.5) { $g$}; 
	\node[below, black] at (-2,-1.5) { $g$}; 
	
	\end{tikzpicture}	
	\caption{The ontic state of the entangled particles is determined by the choices of measurement directions (so statistical independence is violated).}
	\label{gt3}
\end{figure}

\begin{figure}
		\centering
		\begin{tikzpicture}[scale=0.7]
		
			\coordinate (mam23) at (-4.2,0.1);
		\coordinate (mbm23) at (3.8,0.1);
	\coordinate (a3) at (-4.2,-1);
		\coordinate (b3) at (3.8,-1);

		\coordinate (s) at (0,0);
			\coordinate (sm) at (0.3,-0.3);
				\coordinate (sp) at (-0.3,-0.3);
			\coordinate (smin) at (0,-0.6);
			\node[below, black] at (s) { $S$}; 
		\coordinate (y) at (2,2);
			\node[below, black] at (y) { $Y$}; 
				\coordinate (yp) at (1.5,1.7); 
					\coordinate (yp2) at (1.5,1.3);
						\coordinate (yp3) at (2.5,1.3); 
			\coordinate (ym) at (2.5,1.7); 
		\coordinate (mb) at (4,1);
		\coordinate (mbm) at (4,0.2);
			\coordinate (mbm2) at (4,0.1);
				\coordinate (mb3) at (3.5,1); 
			\node[below, black] at (mb) { $M_b$}; 
		\coordinate (b) at (4,-1); 
			\node[below, black] at (b) { $B$}; 
		\coordinate (l) at (0,-2); 
	\node[below, black] at (l) { $\lambda$}; 
				\coordinate (ma) at (-4,1); 
				\coordinate (mam) at (-4,0.2);
					\coordinate (mam2) at (-4,0.1);
						\coordinate (ma3) at (-3.5,1); 
		\node[below, black] at (ma) { $M_a$}; 
			\coordinate (x) at (-2,2); 
			\coordinate (xp) at (-1.5,1.7); 
			\coordinate (xp2) at (-1.5,1.3); 
				\coordinate (xm) at (-2.5,1.7); 
				\coordinate (xp3) at (-2.5,1.3); 
		\node[below, black] at (x) { $X$}; 
		\coordinate (a) at (-4,-1); 
		\node[below, black] at (a) { $A$};

					\draw[black, arrows={-triangle 90}] (l) -- (mam);
		\draw[black, arrows={-triangle 90}] (l) -> (smin);
		\draw[black, arrows={-triangle 90}] (l) -- (mbm);
		\draw[black, arrows={-triangle 90}] (s) -- (yp);
		\draw[black, arrows={-triangle 90}] (s) -- (xp);
		\draw[black, arrows={-triangle 90}] (ma) -- (xm);
		\draw[black, arrows={-triangle 90}] (mb) -- (ym);
		
		\draw[black, dotted, arrows={-triangle 90}] (sm) -- (yp2);
		\draw[black, dotted, arrows={-triangle 90}] (sp) -- (xp2);

 	\draw[black, dotted, arrows={-triangle 90}] (a) -- (mam2);

		\draw[black, dotted, arrows={-triangle 90}] (b) -- (mbm2);
			\draw[black, dotted, arrows={-triangle 90}] (ma3) -- (xp3);
				\draw[black, dotted, arrows={-triangle 90}] (mb3) -- (yp3);

					\node[below, black] at (2,-1) { $g$}; 
						\node[below, black] at (-2,-1) { $g$}; 

 	\draw[black,   arrows={-triangle 90}] (mam23) -- (a3) ;
	
	\draw[black,   arrows={-triangle 90}] (mbm23) -- (b3);

		\end{tikzpicture}	
		\caption{Both the ontic state of the entangled particles and the actual  measurement settings are determined by some hidden variable $\lambda$   (so statistical independence is violated), and the choices of measurement directions are then determined by the measurement settings - in effect the hidden variable is responsible for the choices of measurement directions, so these choices are not `free' as we typically imagine.}
		\label{gt4}
	\end{figure}
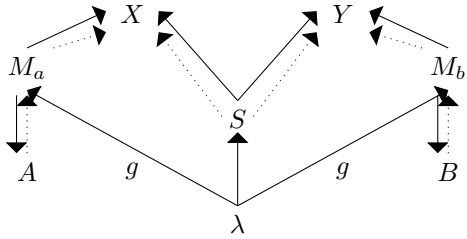

For example, in figures \ref{gt1},\ref{gt3},\ref{gt4} we show DAGs associated with several different generative programs which could be used to generate the empirical data associated with the Bell correlations - in these figures $P$ represents the preparation of the Bell pair and $S$ the true ontic state of the pair after the preparation, $A$ and $B$ represent choices of measurement directions, $M_a$ and $M_b$ represent the actual measurement settings, and $X$ and $Y$ represent the outcomes of the two measurements. Here and throughout the paper we adopt the convention that the `direction of time' as perceived by human observers runs from the bottom of the page towards the top; this representational choice is not part of the definition of the DAG but it makes the graphs easier to read. We use dark arrows to indicate the flow of information prescribed by the generative program, and we include additional dotted arrows showing the causal relations that would appear in standard `common sense' causal accounts of these scenarios, so given our convention for representing time the dotted arrows will always point upwards. It can be seen that generative program DAGs may in some cases have some of the same arrows as the standard causal account, but they can also have arrows going in different directions or connecting different events.

 The nodes of the DAG can loosely be understood as `variables' (or sets of variables) in the sense that they can take a range of values. 
We will take it that a properly-specified generative program specifies clearly what variables are associated with each node and what possible range of values they have: so the `constrast class' for each variable is determined by the program and we do not have to further specify a contrast class when interrogating features of the model. However, our intention is to be quite permissive about what kinds of variables might compose a node -  nodes correspond to entities which are postulated to play an  active role in generating empirical data, so our framework allows people with different metaphysical views about what kinds of entities there are to make different choices about the entities they place at the nodes of their DAGs.

Commonly when DAGs are used to encode causal models they are based on an interventionist understanding of causation and thus the nodes represent the kinds of observable events on which  agents could intervene.  But in our framework we will not insist that the nodes must be things on which observers can intervene: a single node could include variables pertaining to events widely distributed across spacetime, or could   include variables describing  unobservable elements of ontology like fundamental particles, or even variables describing things which are not in spacetime at all, so for example one could imagine a node corresponding to the value of some constant of nature.
Indeed, for those who adopt a robustly realist attitude to laws of nature such that  laws are understood as playing an active role in producing the content of reality, it might sometimes be appropriate to have a node of the DAG corresponding to a law, thus  making explicit the role of the law in the generative process. (Our framework, however, does not make any assumptions about the nature of lawhood: those who do not think of laws in this way can also represent their views in the framework, since they can simply refrain from postulating DAGs which include laws as nodes). As long as the program specifies clearly that these things are variables and tells us what range of values they can take, they are all legitimate nodes in the DAG. 

 For a given DAG it may not always be possible to distinguish cleanly between  nodes corresponding to variables which are observable and nodes corresponding to variables which are not unobservable - as demonstrated by the opponents of logical positivism and constructive empiricist, this distinction cannot usually be drawn in a completely precise way \cite{CONTESSA2006454,Churchland1985TheOS}. And indeed, no part of our framework depends on the claim that DAGs can be neatly divided into observable and unobservable parts; in general we will be interested in DAGs which have substructures that can be recognised as corresponding to some of the empirical predictions of quantum mechanics (e.g. nodes corresponding to quantum measurement outcomes) but beyond this we will not need to specify which nodes feature  directly in conscious experience and which do not. 
 
 There is also a further question about whether conscious experience supervenes only on (some subset of) the nodes of the DAG, or whether observers can be understood as having some empirical access to the structure of the DAG as well - for example, those who espouse 
what we will refer to as the `direct-access' approach to the flow of time believe that our perception of time flowing is the result of some kind of direct empirical access of the process of `temporal becoming' in which the future is generated from the past \cite{10.1093/acprof:oso/9780199218219.003.0005}, so direct-access theorists will presumably want to say that conscious experience supervenes on the arrows as well as the nodes of our DAGs. By contrast, those who espouse  a `perspectival' approach to the flow of time believe that our experience of the flow of time is some kind of illusion arising from the nature of our perspective on reality \cite{articleGodel,Price2005-PRICP}, and they will presumably want to say that conscious experiences supervenes only on the nodes.
Our framework can be used by either direct-access theorists or proponents of the perspectival approach - the two groups will simply draw different conclusions about which DAGs are empirically equivalent.

We will assume that for each generative program there exists a `maximally fine-grained' version of the corresponding DAG, i.e. the graph contains as many distinct nodes as allowed by the generative program. Specifically, this means that variables should be separated into distinct nodes if the program calculates them in two distinct steps. For example,  one might imagine that a generative program for the Bell correlations generates first one measurement result and then the second (possibly conditional on the result of the first), so the measurement results can be treated as two separate nodes; but some generative programs for the Bell correlations might instead generate the two measurement results  \emph{jointly} conditional on the two measurement directions, so the measurement results cannot be treated as two separate nodes. The example DAGs we show in this article will not always be maximally fine-grained, as sometimes it is easier to see the overall graph structure if we combine several nodes into a single composite node, but it is important to keep in mind that the maximally fine-grained version is the one which should be employed when we are making judgements about properties of the program like locality and superdeterminism (see section \ref{ncs}).

\subsection{Original Vertices}

Usually a generative program will require some inputs which do not depend on anything else within the program, which we will refer to as `original vertices.' For example, in traditional deterministic dynamical theories the initial conditions are `original vertices,' as are the time-evolution laws, and the generative program determines the remaining course of history entirely from those original vertices. Similarly, in operational theories the agent is usually regarded as external to the physical theory and thus in an operational theory the decisions made by agents about actions like state preparations and measurements are `original vertices.' On the other hand, physical theories with a more realist flavour may aspire to model agents within the theory, and therefore they may not treat the decisions made by agents as original vertices: instead an agent may be regarded as a physical system which takes inputs (such as sensory information and memories of the past) and produces outputs according to some internal process which can in principle be predicted within the physical theory.  

By treating a vertex as an original vertex we are adopting an official attitude of agnosticism towards its origins, i.e. we are stipulating that its value is simply to be taken as given and there is nothing more to be said about where that value comes from.  So for example, one should not imagine that the value of an original vertex $v$ is selected in a probabilistic way - that would involve postulating an objective probability distribution and  measure outside of the generating program, but by stipulation if such a thing is known it should be included in the generating program, so a variable of this kind would not be original. However, of course we may still assign \emph{epistemic} probability distributions over the values of the original vertices, provided we are clear that these distributions represent our own uncertainty about their values rather than any objective probabilistic generating process. And since   there are by definition no known facts which could play a role in determining the value of $v$, we may invoke the principle of indifference to say that any agent making use of the generating program to arrive at predictions should assign a uniform epistemic probability distribution over the possible values of the original vertex. Here we stipulate that original vertices should always be defined in such a way that a uniform epistemic probability distribution over the values of the variables at the original vertex will induce epistemic probabilities downstream which match the predictions of the theory, ensuring empirical adequacy - in this context `uniform' means simply `uniform according to the natural Lebesgue measure over the reals,' but   this is merely a convenient choice, since one could use any other range and measure provided that subsequent processing steps in the generating program are suitably adjusted to ensure the program predicts the expected probabilities. 

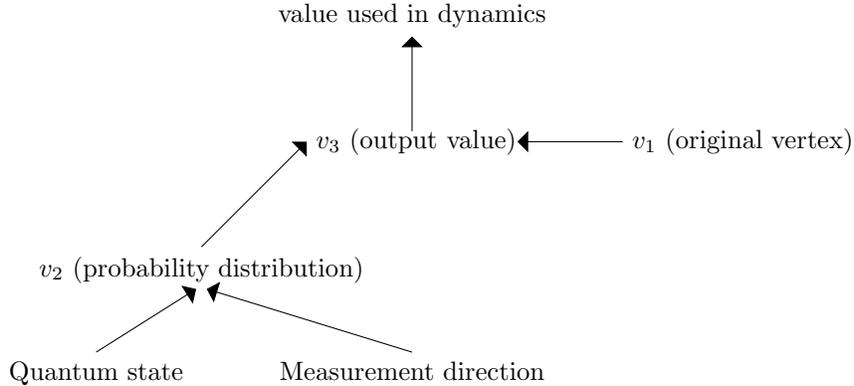
\begin{figure}
		\centering
		\begin{tikzpicture}[scale=0.7]

		\coordinate (past) at (0,0);
		
			\coordinate (consciousness) at (2,2);
			\coordinate (consciousnessplus) at (2.5,2);
			
				\coordinate (action) at (4,4);
				
				\coordinate (rov) at (4,2);
    \coordinate (rov2) at (8,2);

		\node[below, black] at (past) {$v_2$ (probability distribution)};  
\node[below, black] at (-2,-2) {Quantum state}; 
	\draw[black,   arrows={-triangle 90}] (-2,-2) -- (-0.1,-0.8);  

\node[below, black] at (4,-2) { Measurement direction}; 
\draw[black,   arrows={-triangle 90}] (4,-2) -- (0.1,-0.8);  
  
			\node[right, black] at (consciousness) {$v_3$ (output value) };  
				\node[above, black] at (action) {value used in dynamics};  
				\node[right, black] at (rov2) { $v_1$ (original vertex)};  
		
		\draw[black,   arrows={-triangle 90}] (4,2.2) -- (action);  
			\draw[black, arrows={-triangle 90}] (past) -- (consciousness);  
				\draw[black,   arrows={-triangle 90}] (rov2) -- (6,2);  
		
		\end{tikzpicture}	
		\caption{DAG modelling a probabilistic quantum measurement with an original vertex.}
		\label{vertex}
	\end{figure}

Of course, there may also be cases where a generative program \emph{does} postulate an objective probability distribution over the value of some variables, so we can also have  probabilistic variables, which are different from original vertices. In this paper we will take it that probabilistic variables can be modelled by an original vertex $v_1$ combined with another vertex $v_2$ representing the probability distribution   specified by the generative program - for example, if the program prescribes that a certain variable has value $a$ with   probability 0.75 and   value $b$ with probability 0.25, we could model this using an original vertex  $v_1$ which can take any value in the range $[0, 1]$,and an additional vertex $v_2$ whose value will here be set to $0.75$, and then we combine the two vertices to give a third vertex according to the prescription: if $v_1 \leq v_2$ then $v_3 = a $, otherwise if $v_1 > v_2$ then $v_3 = b$.  Because $v_1$ is an original vertex, no objective probability distribution can be postulated over it, but as we have just specified,  users of the program will associate it with a subjective probability distribution, which by convention we are taking to be a uniform distribution over the range $[0,1]$, and thus this program will reproduce the desired probability distribution over $a$ and $b$.  Note that usually vertex $v_2$ will have other arrows going into it representing the prior factors which determine the probability distribution (here fully specified by just setting the value $v_2 = 0.75$)  - for example, if the probabilistic event is a quantum measurement, $v_2$ will have arrows going into it from  the quantum state and measurement direction which together determine the probability distribution for the measurement outcomes - see figure \ref{vertex}.

Throughout this article we will largely deal with generative programs which do not represent a whole universe but instead produce a complete physical description of some particular experimental scenario. In such cases we may have background knowledge that the original vertices are not `really' original, since their values are fixed by processes in the world outside of the experimental scenario. But it is still appropriate to regard them as `original' relative to the physical description of the experimental scenario because they   depend only on parts of the universe not included in the  scenario, so in the context of this specific scenario there is indeed nothing more to be said about where they come from. In the standard terminology of the causal modelling framework, the original vertices are `exogenous' variables which are assumed not to depend on events taking place within the experimental scenario being studied.  An experimental scenario or subregion of spacetime which can be given a self-contained description conditional on a set of variables treated as exogenous can essentially be regarded as something like an independent universe, with the original vertices playing the role of a `boundary' for the universe and the dependent vertices making up the bulk.

\subsection{Ontological Priority}

 Although all physical theories must provide some generative program in order to come up with some empirical data, there are different possible views that one might hold on the significance of these programs.
 For example, operationalists and constructive empiricists \cite{van1980scientific} do not in general commit to any ontology beyond the observable, so they will  presumably not attribute any ontological significance to the generative programs they employ to arrive at empirical data - they will simply regard a generative program as a convenient codification of certain observable regularities. Realists, on the other hand, are committed to the existence of entities or structures which are not observable, and thus   they may be inclined to attribute ontological significance to their generative programs. In particular, we suggest that in the most general case realists may  think of generative programs as encoding information about \emph{ontological priority}. ( \cite{sep-dependence-ontological,Hildebrand2020-HILPLO}). 
 
 Here we say that `A is ontologically prior to B,' just in case B depends on A for its existence - for example, if one believes that the universe unfolds by a continuous evolution process involving `dynamical production,' then one will naturally think that later states of the universe depend on earlier states of the universe for their existence, so earlier states are ontologically prior to later ones.   In principle all sorts of things may depend on other things for their existence - for example, one might think that `the fact that A \& B' depends for its existence on `the fact that A' and `the fact that B.' However, in this article we are concerned specifically with ontological priority construed as a physical relation between distinct real entities, so we will not consider applications of it outside of the physical domain. 
 
 Ontological priority is often described metaphorically as representing the order in which god would have to create things in order to arrive at the universe - or in less theological terms, one could think of it as a computer program specifying the steps one would have to take in order to arrive at a complete universe \cite{Hildebrand2020-HILPLO,Schaffer2009-SCHOWG}. This metaphor makes it clear why it seems natural to interpret `generative programs' in terms of ontological priority. Ontological priority is a reasonably common concept in philosophy, but has not yet made its way into physics in any explicit way, perhaps because physicists tend to be sceptical of notions which sound excessively `metaphysical.' But really ontological priority (at least insofar as that term is understood as pertaining specifically to nomic or modal relations between distinct real entities) is not any more metaphysical than `causation,' so in our view, anyone who is willing to accept `causation' as falling in the domain of physics should also be willing to accept `ontological priority.' After all, causation can be understood as simply a   special case of ontological priority: if one thinks of the statement `A is the cause of B' as  an assertion that B depends  for its existence on A, then this statement means that A is ontologically prior to B, and thus standard causal-dynamical physical theories would typically be associated with a generative program in which  the  initial condition of the universe  is  ontologically prior to the rest of history, which is generated from it by time evolution.

On the other hand, as shown by the diagrams in section \ref{programs}, relations of ontological priority don't have to coincide with our presuppositions about causal structure - for example, we typically imagine that causation goes forward in time, but we can certainly write down programs in which future events are ontologically prior to past ones. In this article we will use the term `strong causation' to refer to relations which are self-evidently causal according to our common-sense understanding of causation (i.e. mostly relations between macroscopic events where the cause precedes the effect in time); there is probably a case to be made that other relations of ontological priority might also qualify as `causal' according to some generalized notion of causality, but we will not use that terminology in this article. 
Now, those who espouse a direct-access approach to the flow of time typically believe that our common-sense notion of strong causation reflects our experience of a real structure of temporal becoming, and thus they may argue that DAGs in which relations of ontological priority disagree with  intuitions about strong causation are not candidates to represent the actual world, since they don't get our experiences of temporal becoming right. However, proponents of the perspectival approach to the flow of time in which experience supervenes only on the  the nodes and not the arrows of a DAG will presumably take the view that a DAG in which relations of ontological priority disagree with  common-sense intuitions about strong causation could be a candidate to represent the actual world. 

Furthermore, relations of ontological priority can also apply to  more general relata than causal relations, including relata which may not be located in spacetime, because relations of ontological priority may hold between any entities which play an active role in the way in which a generative program produces data, regardless of whether or not the associated theory is usually interpreted as saying that entities of this kind live in spacetime. For example, fundamental constants,  mathematical objects like Lagrangians, or even laws of nature could stand in relations of ontological priority, even though these kinds of objects are arguably best understood as non-spatiotemporal.

 \subsection{Generalizing Causal Modelling \label{general}}

There is a well-recognised connection between causation and ontological priority. Schaffer, for example, has noted that there is a close connection between causation and a philosphical concept known as `grounding,' which is  related to ontological priority:   `\emph{the analogies run deep: both (causation and grounding) feel like relations of generation, both look something like partial orders, and both can back explanation}' \cite{Schaffer2016-SCHGIT-2}. Schaffer therefore argues that grounding can be represented by a system of structural equations, akin to the structural equations used to represent causation in causal modelling, and thus the same mathematical formalism can be used to understand them both.\footnote{Our approach has much in common with Schaffer's, but ultimately has different subject matter. In particular, our definition of ontological priority is both narrower and broader than  the usual definition of grounding:  on the one hand, grounding is usually understood as subsuming all kinds of cases where one entity depends on another for its existence, so for example one might say   ‘the fact that A \& B’ is grounded by ‘the fact that A’ and ‘the fact that B,' whereas we have specified that ontological priority applies only to physical relations between distinct real entities. But on the other hand, our definition  entails that ontological priority includes causation as a special case, whereas Schaffer at least does not see causation as a special case of grounding. In addition Schaffer's  primary motivation is to better understand the concept of grounding (via a non-reductive analysis), whereas our focus here is on showing how ontological priority can be applied to real scientific problems and arguing that it can be a scientifically useful tool.   This means we   do not have to deal with concerns like that raised by Jansson \cite{Jansson2018WhenAS}, who contends that using structural equations to make inferences about grounding relations is problematic because we don't have access to the kinds of a posteriori knowledge  needed to decide which variables the DAG should include or what the appropriate contrast class for each of the variables is - this issue does not arise in our case, because   we are not attempting to use structural equations to make inferences about relations of ontological priority, rather we simply take it that  the generative program  specified within a theory stipulates what the relevant ontological variables are and what possible alternative values they have, so the purpose of the DAG is to clarify the nature of these commitments rather than to figure out what they are.}

Here we follow Schaffer in arguing that because   ontological priority is naturally represented by a directed acyclic graph, we may apply some mathematical techniques developed in the study of DAGS representing causal models (e.g. see refs  \cite{pearl2009causality} and   \cite{spirtes2000causation}) to our own DAGs encoding ontological priority. For example, as in Schaffer's grounding case, we can use generative programs to investigate how relations of ontological priority ground counterfactual relations - the   generative program prescribes  relations of functional dependence between nodes, so we can make quantitative statements about the outcome of making changes to the value of a variable at one node of the graph while leaving other inputs to the program unchanged, and moreover we can do this regardless of whether or not the node in question is one on which realistic agents could actually intervene.  For example, suppose we have a generative program where the value of some constant of nature (e.g. the cosmological constant) is an original vertex, and this vertex is used to determine the state $S$ of the world at some time $t$. Of course no real agent can intervene to vary the cosmological constant, but nonetheless we can use the generative program to imagine   `intervening' by changing the value of the cosmological constant whilst keeping the program and the values of the other original vertices  constant: if the program is properly formulated we should be able to calculate a new state $S'$ for the world at  time $t$, so we  can check whether it is the case that, according to this particular generative program, the state $S$  counterfactually  depends on the value of the cosmological constant. The practice of studying the consequences of varying fundamental constants is actually very common in physics (see refs  \cite{Hogan_2000,Tegmark_1998}) and our notion of ontological priorty can be understood as formalizing the reasoning behind this practice.

To make mathematical statements about the dependence relations encoded in a generative program, we will make use of   a probability distribution over original vertices (although as already noted, this distribution should be understood epistemically rather than ontologically).  For example, we may wish to stipulate that part of what it is for a generative program DAG to be a correct representation of a possible generative program is that that program should satisfy the causal Markov condition relative to that DAG - i.e. when we consider the outputs of the generative program given independent random choices of values at the original vertices, we should find that for each vertex $v$, conditional on the set of its `parents' in the graph (i.e. all vertices  $x$ such that there is an arrow from $x$ to $v$ ) $v$ is statistically independent of all variables which are not its parents or its descendants \cite{pearl2009causality}. Of course in practice the original vertices only take one value so the causal Markov condition can't be verified using any actual empirical data, but nonetheless we can use this condition to check whether or not a proposed DAG  is  consistent with the generative program it is supposed to represent - if the program generates variables which fail to exhibit the right independence relations, this suggests that the generative program involves some further relations of ontological priority that we have failed to capture in the DAG. 

Similarly,   in causal modelling it is common to require that causal graphs should exhibit `faithfulness,'  i.e., when we consider independent random choices at the original vertices, two variables in the graph should be statistically independent only if they are `d-separated' in the underlying graph, which is to say that any paths between them are `blocked' by other variables playing a role somewhat like a common cause \cite{pearl2009causality}. So in particular, a graph fails to exhibit faithfulness if there is a causal arrow from A to B but the distribution over values obtained from independent random inputs at the original vertices shows A as being statistically independent of B. And similarly, it seems natural to require that graphs of ontological priority should similarly obey faithfulness, for if there are unexpected independences in the empirical data, this may indicate that there is an opportunity to add something further to the generative program that has been proposed - we discuss this point further in section \ref{finetuning}.

\section{Realism versus operationalism}

We now turn to demonstrating some useful applications of the generative program framework. First, we use it to make precise the nature of the commitments associated with `operationalism' and `realism,' Roughly speaking,   operationalists are those who regard physical theories as simply providing   succinctly-expressed generalizations about empirical data: they do not purport to \emph{explain} the data or say how it comes about. Whereas `realists'   want to do more than this - they aspire to produce theories which do actually explain the empirical data. However, opinions diverge on what exactly that means, and thus there are many different forms of realism. One possibility is     what one might refer to as `object-oriented realism.' Realists of this type usually suggest that realism involves explaining empirical data by postulating a set of objects or entities whose behaviour is responsible for the observed effects - for example, the `primitive ontology' approach to quantum mechanics argues that we will not understand quantum mechanics until we have expressed it in terms of ontological building blocks living in three-dimensional space ( \cite{Esfeld_2020,pittphilsci11651}). Another possibility is what is commonly referred to as `structural realism.' Realists of this type take it that the essence of realism is explaining empirical data by appeal to some underlying objective structure - and commonly it is assumed that this structure is supposed to be a causal one. For example, proponents of the causal modelling programme for quantum mechanics argue that we must separate out elements of the theory which are merely epistemic from the underlying causal structure, which they identify with the `realities of nature' ( \cite{esfeld2009modal,schmid2020unscrambling}).  

However, in our view  both of these forms of realism are too narrow. The object-oriented approach is problematic because  our access to unobseravble entities or objects is limited and indirect, particularly at the small scales with which many parts of modern physics are concerned. As structural realists have been urging for some time ( \cite{Ainsworth2010-AINWIO-2,Berenstain2012-BEROSR}), our data tells us most directly about \emph{structures} and there is no guarantee that our attempts to infer ontology from those structures will succeed - for all we know, the true underlying ontology could be so radically different from anything in our experience that we are not even capable of conceptualizing it. The structural approach avoids this issue by focusing specifically on structure, but when it is couched in terms of causal structure it is also flawed,   because this approach takes for granted that reality must have a causal structure at the most fundamental level, and we have seen that there are good reasons to think that may not be the case. This leaves us with a dilemma - what does it mean to be a realist if one does not espouse object-oriented or causal accounts of fundamental physics? If fundamental physics does not have a causal structure, does that mean it is hopeless for us to make any attempt to explain empirical data? Surely not - causation cannot be the only kind of structure which could possibly play an explanatory role. But what kind of structure could replace it?

The literature contains various suggestions for approaches to non-causal explanation which might help fill this gap. However, in many cases the `realist' credentials of these approaches seem somewhat unclear. For example, Felline \cite{e23050589} proposes explaining the Bell correlations by appeal to structures which are not merely higher-level descriptions of more fundamental entities and processes but which are `mechanistically fundamental.' This seems like a promising starting point, but since there are a large number of different structures featuring in scientific theories,  more needs to be said about what kinds of structures can be `mechanistically fundamental' and what realist commitments are justified by these structural explanations. The term `fundamental' does not necessarily help here, because ideas around fundamentality in physics are often quite vague and different people have different intuitions about what counts as fundamental. 

But the generative program framework offers a useful way of formalising non-causal structures. Specifically, we suggest that   it is the structures encoded in generative programs, i.e. structures encoding ontological priority, which should replace causal structure to give a more general version of causal structural realism; structural realists can be committed to the   ontological priority structure encoded in a generative program DAG in just the same way as they may be committed to the causal structure encoded in a causal DAG, and thus this framework provides a locus for structural realist commitments even in domains where a purely causal description is inappropriate. 

Of course,  both operationalists and realists may make use of generative programs, so to clarify the nature of the realist's commitments we need to understand the ways in which realists and operationalists differ in their use of generative programs. To begin, we  note that for most kinds of empirical data there will be many different generative programs which could predict the same empirical data. For example,   suppose  we observe that whenever an event of type A occurs, then  event B follows with certainty, but B does not occur unless A has occurred just prior. It would seem natural to describe this phenomenon by a generative program such that whenever we generate an event of type A we then generate an event of type B at a later time, which  would be represented by a  DAG with an arrow going from the vertex associated with A to the vertex associated with B. However, one can also define a consistent generative program in which whenever we generate an event of type B we then generate an event of type A at an earlier time, which  would be represented by a  DAG with an arrow going from the vertex associated with B to the vertex associated with A. The latter possibility involves something that looks a little like `retrocausality,'  though the relation going backwards does not qualify as strong causation since our common-sense judgements of causation usually assume causes precede effects in time. Thus this possibility does not accord with the way we would typically narrativize this set of events, but at least for those who espouse a perspectival view of the flow of time it is still a viable program which generates the right empirical data. So we suggest the difference between realist and operationalists lies in the way they will interpret these kinds of empirically equivalent alternative programs:  operationalists do not regard physical theories as having representative content or as asserting anything about reality beyond the described regularities, and therefore they cannot say that there is a fact of the matter about which entities stand in relations of ontological priority to each other, so they must take it that empirically equivalent generative programs are simply equivalent representations of the regularities observed by agents. On the other hand, realists about scientific theories typically believe that reality does have some objective structure of its own, and  thus they  may regard a scientific theory as putting forward a hypothesis to the effect that some particular generative program DAG is a correct representation of the actual structure of our reality. Thus for the realist, generative program DAGs are not merely recipes for arriving at empirical data, but constitute substantive claims about the structures that generate that data - so from the realist point of view generative programs can be right or wrong in what they say about ontological priority, whereas from the operationalist point of view it would be nonsensical to describe an empirically adequate program as right or wrong.

    That said, we emphasize that the realist need not be committed to the claim that \emph{any particular} DAG is the correct one - it is possible to be a realist whilst also maintaining a position of epistemic humility which emphasizes the fallibility of scientific knowledge. So the prudent realist will accept that we cannot know for sure that the generative program specified by a theory matches the true structure of reality, while maintaining that there is nonetheless a fact of the matter about which generative program best captures the  structure of reality.  We also emphasize that realists need not regard any two superficially distinct programs as being ontologically distinct - for example, most realists would probably consider that two different programs for classical electromagnetism which use different gauge fixing schemes correspond to the same generative program, since different configurations of the potentials related by a gauge transformation are typically taken to represent the same physical situation. So deciding which programs count as distinct and which parts of those programs should be reified will always be a nontrivial problem; but the fact remains that realists must believe that there is some real, ontological structure which explains why we get the data we do, and therefore  once we have decided that two programs are distinct it follows that one may be better than the other at  capturing this structure. 

  We can apply our definition to answer the perennial question of whether or not QBism is really a form of realism. QBism is an interpretation of quantum mechanics which asserts  that quantum mechanics does not predict or summarize empirical data: rather it is a normative principle which tells rational agents how they should form their beliefs \cite{QBismintro,2010arXiv1003.5209F}. Thus according to QBism, quantum mechanics is not a physical theory in the sense in which we have used that term, and thus it should not be associated with a generative program or DAG at all since it does not  describe empirical data so there is nothing for the program to predict. Evidently if this is the end of the story QBism cannot be a form of realism, since it does not postulate any DAGs at all so it certainly cannot involve a commitment to any particular DAG. However, some QBists do want to identify themselves as realists - they believe in an external world which is responsible for our experiences, they simply don't think \emph{quantum mechanics} can be regarded as representing anything about that external world \cite{pittphilsci16382,fuchs2016participatory}. So this   class of QBists would presumably accept that there does exist some generative program and DAG representing the structure of the external world which is responsible for our experiences, while denying that this DAG has anything to do with with quantum mechanics.  And this class of QBists do indeed  count as realists according to our classification, since we did emphasize that realists need not be committed to the claim that we actually know the DAG which represents the structure of our actual world - though it might be argued that this group of QBists are taking epistemic humility too far by refusing to even contemplate what the DAG responsible for the empirical data could look like.

\subsection{Intersubjectivity and Persistence Over Time \label{intersubjectivity}}

We can also apply the generative program framework to bring out some important consequences of taking an operational approach. We have argued that the founding tenet of  operationalism is that  all `empirically equivalent' physical programs are equally good; evidently  this means that the operationalist must be able to specify clearly which parts of a graph of ontological priority are to be regarded as empirical and which are not. As we have already noted, this is in general a non-trivial thing to do. Typically, operationalists seek to make this distinction by postulating an  in-principle split between the agent and the rest of the world, so we are to imagine an external agent acting on reality and getting back certain responses, which are the data to be predicted by the theory\footnote{See refs  \cite{PhysRevA.81.062348,abramsky2013operational,hardy2016operational} for some examples of operational formulations of theories. However, we emphasize that the authors of these papers may not necessarily identify as `operationalists' - it is possible to adopt an operational approach to study the structure of theories without necessarily rejecting the possibility of an underlying realist story.}. This allows the operationalist to identify what counts as `empirical' in a somewhat more clean way - in particular, they don't have to worry about which parts of the internal structure of the agent's decision-making process should count as empirical. Evidently operationalists of this kind   are restricted to choosing generative programs which represent their own choices as original vertices, although they will not be committed to any definite structure for the rest of the graph.   

But what happens if we have two or more different agents? In that case, there are two options: either we adopt a first-person operationalism in which each agent aims to construct a generative program to reproduce only their own observations, or we adopt a collective operationalism in which the aim is to construct a generative program to reproduce the observations of all of the agents simultaneously. However, the latter approach seems quite hard to justify within an operationalist picture, because of course `the experiences of other agents' are not directly observable or accessible any more than the other unobservable entities that the operationalist doesn't want to commit to, so  an operationalist who is committed to the reality of the observations made by other agents but not to any other unobservable entities seems to be making a somewhat arbitrary distinction. Thus, in fact operationalism is usually understood in a first-person way (this has been advocated explicitly in the context of pragmatic and operationally-inclined interpretations of quantum mechanics, as in refs  \cite{Cavalcanti_2021,wiseman2015causarum}), which is in accordance  with the operationalist maxim that a scientific theory is just a tool that an agent can employ to predict the measurement results that they will witness. 

What this means is that if Alice is an operationalist, her generative program will usually represents her own decisions as original vertices, but it need not represent  Bob's decisions as original vertices: he is just another physical system which she can perform measurements on and which may or may not include some original vertices. Furthermore, in the first-person approach Alice is committed to the view that all programs which predict the same empirical data \emph{for her} are equally good, and therefore she cannot attach any special significance to programs which represent Bob's decisions as original vertices: from the operationalist point of the view there is no fact of the matter about whether they are original vertices or not.  So Alice necessarily has a description of the world which is different from Bob's, because both of them give an account of reality which involves quotienting over a set of programs, and only a small subset of Alice's programs will be in Bob's set and vice versa.  Thus  the operationalist view is not   compatible with the existence of a single  generative program which  generates the entire universe: rather operationalists must employ a different set of generative programs $p_A$ for each agent $A$, such that all of the programs in $p_A$  represent the actions of $A$ as a set of original vertices but the programs in $p_A$ do not as a general rule represent the actions of other agents as a set of original vertices.

The differences between the sets of programs associated with distinct agents are even more pronounced in operational approaches to quantum mechanics. For operationalists typically hold that an agent should apply unitary quantum mechanics universally to the external physical world, and when this prescription is applied in situations like a Wigner's friend experiment \cite{pittphilsci19401}, it has the consequence that different agents can disagree not only on the \emph{structure} of their generative program DAGs but also on the empirical data generated by the generative programs: Wigner's friend will use a generative program which specifies a unique outcome for his measurement, whereas Wigner will use a generative program which says that the measuring device remains in a coherent superposition so there is no measurement outcome at all. The Frauchiger-Renner experiment makes this even more explicit by showing that in a certain case the programs used by different agents will predict incompatible outcomes to one and the same measurement \cite{2018qtcc}.  This case, and the related Local Friendliness arguments \cite{Bong_2020,utrerasalarcón2023allowing} also highlight that while an agent chooses inputs and receives outputs which become part of a permanent record,  they model other agents as having no permanent record of their experiences; so in the quantum context, we end up with a set of   disconnected generative programs used by different agents, and though  these sets of generative programs will display roughly similar statistical features they nonetheless represent  different physical worlds with both different structure and also different empirical content. Thus in general operational theories cannot provide us with any unified account of an intersubjective reality shared by different agents;  in addition to being counterintuitive, this also raises serious questions about the scientific method and empirical confirmation in the context of operational theories (see  ref  \cite{https://doi.org/10.48550/arxiv.2203.16278} for a discussion of the importance of intersubjectivity  in empirical confirmation).

These problems with intersubjectivity also extend to intersubjective agreement between different versions of the same observer at different times - for in an operational   context it is   difficult to  model agents persisting over time without undermining the basic premises of operationalism. After all, part of what it is to be an agent persisting over time is to retain memories and act in ways that are partly influenced by one's past experiences, and this surely requires that the agent is involved in some kind of two-way interaction with the systems described by the theory:  the external world must have something to do with the memories and experiences that define the stable character of the persisting agent.  One might expect this   to give rise to a generative program in which  measurement outcomes in the past are used to generate an agent's future choices, or   measurement outcomes and future choices are generated jointly by the program as in an all-at-once model.  But using a program of this kind  amounts to accepting that the agent is not, after all, separate from the physics: at least some parts of the agent's decision-making processes are in fact determined within the generative program generating the empirical data. So if we allow some kind of two-way interaction between agents and their surroundings, it becomes difficult to maintain the operationalist's strict separation between agent and world, thus undermining the operationalist's identification of what counts as `empirical' and making it unclear exactly which generative programs we are supposed to be quotienting over.

The generative program framework is useful here because it  compels the operationalist to make a choice about what their position is really saying about agency: either they must use a completely distinct generative program to predict the result of each individual experimental situation, leaving us with an impoverished and fragmented approach, or they must accept that the   agent is, in at least some rudimentary way, modelled as part of the generative program, meaning that we are no longer adhering to a strict operationalist position. An example of such an intermediate possibility is given by   ref  \cite{https://doi.org/10.48550/arxiv.2203.13342}, which is a version of relational quantum mechanics with an additional postulate aiming to secure the possibility of intersubjective agreement between different agents.  In this approach, facts are regarded as being relative to `observers' (where every physical system can be an observer) and thus each observer, including different versions of the same observer at different times, has their own perspective on reality such that these perspectives may sometimes contain incompatible empirical facts. But the added postulate  of `cross-perspective links' ensures that agents can come to share facts during certain kinds of interactions, meaning that their perspectives come to agree on at least certain kinds of facts. Thus the  generative program DAG for this theory would have some disconnected parts but would also contain connections between the DAGs associated with different observers, with those connections generated in the course of appropriate physical interactions. It would be interesting to see this idea made more formal - indeed, trying to come up with a well-defined generative program DAG for this version of RQM could be an important test of its coherence, because this process would require us to resolve  some open questions about how exactly the connections in this picture can be constructed.

\section{Nonlocality, Contextuality and Superdeterminism \label{ncs}}

We may now apply the distinction between realism and operationalism to help clarify a somewhat contested issue in physics - what exactly does it mean to give a realist account of the Bell correlations? In ref  \cite{schmid2020unscrambling}, for example, it is suggested that a realist account must be a causal one, but evidently this answer is not much help if in fact the Bell correlations don't have a causal structure, as argued in refs  \cite{e23050589,https://doi.org/10.48550/arxiv.2208.02721}. However, the generative program framework provides a straightforward way to resolve this question: \emph{any} generative program which generates the empirical data exhibited in the Bell experiments can be understood as `realist' provided we are committed to the view that the program corresponds to the true ontological priority structure of reality, or we are at least willing   to entertain the possibility that it \emph{could} correspond to the true ontological priority structure of reality. So programs in and of themselves are not realist or non-realist: rather it is our epistemic attitude towards the program which determines whether we are taking a realist position or not.

This has important consequences for our understanding of nonlocality. There are many different ways of defining the term `locality,' but for our purposes it will be most useful to distinguish two important classes of definitions: some definitions suggest that a theory is `local'  only if it explicitly describes local mechanisms by which the empirical data is generated, and other definitions suggest that a theory is `local' provided that it does not postulate any \emph{non-local} mechanisms (e.g. an instantaneous wavefunction collapse) by which empirical data is generated. While people are free to use words in whatever way they wish, our own view is that the former approach is more useful, because it is neither difficult nor particularly illuminating to come up with a so-called `local' model by simply refusing to give any account of the way in which the data is produced (for example, by being an operationalist or moving to a principle theory description \cite{Koberinski}). This also accords with Bell's own definition: he understands locality as involving the postulation of an underlying mechanism involving only dependence on a variable $\lambda$ fixed at the time of the initial state preparation. So we prefer to say that a theory is local if it explicitly describes local mechanisms by which the empirical data is generated, non-local if it explicitly describes non-local mechanisms by which the empirical data is generated, and neither local nor non-local if it refrains from describing any mechanisms by which the empirical data is generated. 

Now, many existing attempts  to define `explicitly local mechanisms,' assume that the mechanisms in question must be 
  \emph{causal} ones, but we  prefer  not do this since our aim is to avoid defining locality in causal terms. But instead we can appeal to  the generative program framework, which allows for a natural generalisation of the causal approach: let us say that a theory is local if it is associated with a  generative program such that  a) in a maximally fine-grained version of the ontological priority graph associated with the program (as defined in section \ref{programs}), no single node in the graph contains variables which are located at a spacelike separation, and b) no node $N$ containing a variable located at some spacetime point $x$ is connected by a relation of ontological priority to a node $N'$ containing a variable located at a spacelike separation from $x$, unless that connection is composed of a series of nodes each containing variables located wholly in spacetime such that each node's variables are in the past or future lightcone of the previous variable in the sequence. To say that a theory is `non-local' in Bell's sense is to say that the theory is associated with a generative program which represents Bell-type experiments in some other way.

 Requirement a) means that according to our definition of locality, theories with `non-separable' objects or processes spread across space-like separation count as non-local. 
Requirement b), meanwhile, is written carefully so as to allow certain kinds of retrocausal theories. For example,  a retrocausal model where the Bell correlations are mediated by a signal that goes into the future and then back into the past will satisfy requirement b), since the connection between the choice of measurement for one Bell event at spacetime point $x$ and the other measurement at a spacetime point $y$ spacelike separated from $x$ goes via a third node corresponding to an event located in the intersection of the future lightcone of $x$ and $y$ where the   signal arrives at some future point  and then begins to go backwards in time. On the other hand, a model in which the connection between the choice of measurement for one Bell event at spacetime point $x$ and the other measurement at a spacetime point $y$ spacelike separated from $x$ goes via a third node which is not located in spacetime (e.g. a modal or nomic connection) does not satisfy requirement b), since the third node is not located wholly in spacetime.   Note that this requirement can be relaxed if we wish to define locality in a way that does not allow for retrocausality to be local: then it is enough to require that  no node containing a variable located at some spacetime point is connected by a relation of ontological priority to a variable located at a spacelike separation from that point, since there is no way to have two spacelike separated events connected by a series of   nodes each containing variables located wholly in spacetime such that each node's variables are in the  \emph{future} lightcone of the previous variable in the sequence.

  Note that  clearly there exist  both generative programs generating the empirical data exhibited in the  Bell experiments which are local in the above sense (e.g. programs derived from a superdeterministic approach) and also generative programs generating the same empirical data which  are non-local in the above sense (e.g. programs derived from a spontaneous collapse model).   Thus, if one wants to take a position on locality in nature, one must take the philosophical position that there is some fact of the matter about what kind of generative program generates the data, which is ultimately a commitment that goes beyond just describing empirical data.  The recent Nobel prize in physics was awarded to experimental tests of Bell's theorem, so evidently this philosophical question is of great importance to physicists, which is a vindication of the need for philosophical argumentation in modern physics.

In particular, we emphasize that an operationalist approach to the Bell correlations cannot involve a commitment to any particular generative program -   the operationalist may find some programs more convenient than others, but ultimately they must say that all empirically equivalent programs are equally correct descriptions of the empirical regularities.  Thus, straightforwardly, an operationalist approach of the Bell correlations cannot be local or non-local in the above sense, because the operationalist  account cannot be committed to local programs over non-local programs or vice versa. So, insofar as locality is to be understood as a claim about \emph{how} data is produced, it is not correct for an operationalist to say  that their approach to  quantum mechanics is `local' just because it does not postulate any explicitly non-local mechanisms - for their approach does not postulate any explicitly local mechanisms either, since they are not committed to any particular generative programs and thus they are not postulating any specific mechanisms at all.  From this point of view, nothing can be said one way or another about whether an operational approach is local until we supplement it with a commitment to some particular generative programs or proper subset of the empirically adequate programs - and once such a commitment is made then the approach becomes  a realist account, since being committed to  a program or subset of programs in this way entails a commitment to the idea that there is some fact of the matter about the objective structure of the world. This indicates that any operationalist who claims to have a local (or nonlocal) account of quantum physics is either a covert realist or is using the term `locality' in some other way that doesn't tell us anything about how the data was produced.

Similar points can be made about the term `contextuality.'  In the language of the generative program framework, to say that a theory is `preparation non-contextual' in Spekkens' sense \cite{Spekkenscontextuality} is to say that it is associated with a generative program such that the steps employed by the program to arrive at empirical data for a given choice of preparation are the same for all operationally equivalent preparations, i.e. all preparations which are usually regarded as preparing the same quantum state,  and to say that a theory is `preparation contextual' in Spekkens' sense is to say that it is associated with a generative program which does not satisfy this criterion. There exist both preparation contextual and preparation non-contextual generative programs which can generate the empirical data for quantum mechanics (for example, we can arrive at non-contextual programs if we allow the program to violate the assumption about how ontic states are composed which  appears in Spekkens' contextuality proof), and thus since the operationalist cannot be committed to one empirically adequate generative program over another, operationalists cannot be committed either to preparation contextual generative programs or to preparation non-contextual generative programs, so operationalist approaches cannot have either of these properties. 

And the same applies to the term `superdeterministic.'  Following ref  \cite{refining}, we define \textit{No Superdeterminism} as the requirement that the response function determining how a system would have responded to measurements other than the ones the observer actually made   is entirely consistent with the observer's past empirical experiences.  We formalize this by defining the list of \textit{empirically plausible interventions} as the set of all interventions the observer believes they can perform on the system, based on their past experiences, and the set of \textit{empirically plausible responses} as the set of all responses consistent with past empirical data the observer has seen in response to past interventions of various types\footnote{To avoid  confusion, we stress that a general intervention need not occur at a single event, and a general response can be nondeterministic, as long as it has a well-defined probability distribution.}.  The  response function defined from the program must allow all empirically plausible interventions as inputs, and all outputs in the range of the function must also be empirically plausible.   Otherwise the response function would entail something about the future intervention by the observer, and this entailment would be a conspiratorial or superdeterministic pre-arrangement.  This definition is the minimum standard to ensure that the description of the system entails nothing about the observer's future intervention.    The underlying idea is that in a superdeterministic program, no observer is led to believe they can make a choice which is in fact physically impossible given that program.

Obviously, this definition rests crucially on a prescription about how to obtain response functions from a generative program. Taking inspiration from the possible-world semantics for counterfactuals, we suggest that response functions are to be obtained as follows: if the observer actually made measurement $M$, then to calculate the output of the response function for a different measurement $M'$, we should take  the maximally fine-grained ontological priority graph, and change one node $N$ in that graph to transform $M$ into $M'$; then ignore everything which is ontologically prior to $N$ and apply the usual prescriptions of the generative program to calculate the values for all nodes  $X$ such that $N$ is ontologically prior to $X$.  So for example, if the program generates events in accordance with the usual temporal order and the result of the measurement node $N$ is a function of the choice of 
measurement and the state of the system, we can simply change $M$ to $M'$ in that function to calculate the output. But for some `superdeterministic' programs, this won't be possible: there will be empirically plausible choices of measurement $M'$ such that the generative program just won't say anything at all about what the outcome is for that choice of measurement, so indeed the observer is led to believe they can make a choice which is in fact physically impossible according to the program. Note that it also won't be possible to arrive at a sensible response function if the generative program is such that the output of the measurement is actually ontologically \emph{prior} to the measurement choice: in that case the program doesn't allow us to calculate how the output would have been different if the input had been different, so again observers don't have the choices that they imagine themselves to have. 

Note it is important that we change entire nodes at once. For example, in an `all-at-once' model where the whole of history is generated atemporally in one go, the whole of history corresponds to a single node, so according to the prescription above we are not allowed to adjust the measurement choice while leaving the rest of history untouched: to assess the response function we have to change to an entirely different course of history in which a different measurement was made. Provided that for every empirically plausible measurement $M'$ there is some course of history which is possible according to the laws of the all-at-once model in which the observer performed $M'$ rather than $M$ at the relevant point, our response function will have outputs for every empirically plausible measurement so this model won't be superdeterministic. On the other hand, this model is non-local, since the node containing the whole course of history will include many spacelike separated pairs of events. Of course, if instead of treating the whole course of history as a single node, one instead calculates a course of history and then carefully `programs' each  individual particle at the beginning of time to have a trajectory compatible with the desired history, this model is no longer non-local since now the particles can all be treated as separate nodes - but such a model will usually be superdeterministic since the generative program won't provide a prescription for how to calculate the changes to the behaviour of the other particles if we alter the behaviour of just one particle.  Each of the trajectories is fully scripted in harmony in the initial state, and then none of the particles actually intervene upon one another as these scripts play out, even though they appear to interact.

\subsection{Fine-Tuning \label{finetuning}} 

The generative program framework can also be used to gain clarity on contested issues surrounding `fine-tuning.' Specifically, we will say that a generative program is `fine-tuned’ if two or more of the original vertices have values which appear to match in a way that is surprising, given that original vertices are supposed to be independent of one another.

A variety of well-known examples of fine-tuning in physics fit this definition. For example, the flatness problem \cite{GIBBONS1987736}: the current density of matter and energy in the universe appears to be very close to the critical value required for a flat universe. Furthermore,  natural mechanisms ensure that deviations from the critical value tend to increase rapidly over time, and therefore the initial density must be chosen very carefully to match the critical density since otherwise it would be very far from the critical value by now.  Thus if we propose a generative program in which the initial density and the critical value are  separate original vertices,  then this generative program is fine-tuned. However, it is now generally believed that the matching can be explained by the mechanism of inflation \cite{PhysRevD.23.347}, which suggests an alternative generative program in which the density is not an original vertex since it is `generated' during a period of inflation which naturally forces it towards the critical density, and therefore we  can get rid of the fine-tuning by moving to a different generative program.  

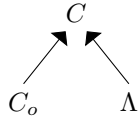
\begin{figure}
	\centering
	\begin{tikzpicture}[scale=0.7]

			\coordinate (cp) at (0.2,0);
			\coordinate (cm) at (-0.2,0);

	\coordinate (c) at (0,0); 
	\coordinate (b) at (-1,-1);  
    \coordinate (bb) at (-0.7,-1.3);  
	\coordinate (a) at (1,-1); 
   \coordinate (aa) at (0.8,-1.3);  
	
	\node[above, black] at (c) { $C$}; 
	\node[below, black] at (b) { $C_o$}; 
	\node[below, black] at (a) { $\Lambda$}; 
	
	\draw[black,   arrows={-triangle 90}] (b) -- (cm); 
	\draw[black,   arrows={-triangle 90}] (a) -- (cp);

	\end{tikzpicture}	
	\caption{Possible structure of ontological priority determining the value of the cosmological constant.}
	\label{cosmofirst}
\end{figure}

Similarly, concerns about `naturalness' in particle physics pertain to the fact that in certain calculations for physical parameters there is supposedly a precise matching between the  bare quantities and the vacuum fluctuation contributions such that they cancel up to many orders of magnitude \cite{Grinbaum_2012}. For example, this occurs in the usual calculation for the cosmological constant - the cosmological constant is very close to zero, and in order to achieve this the contribution due to vacuum fluctuations, which is proportional to the fourth power of the cut-off scale, must very exactly cancel the `bare' value of the cosmological constant \cite{https://doi.org/10.48550/arxiv.2112.12235}. Moreover it seems quite natural to think that both the bare cosmological constant and the cutoff length should be original vertices, as depicted in figure \ref{cosmofirst}, and yet if we propose a generative program in which they are original vertices it will be fine-tuned.

Now, we should clarify what we mean by the claim that original vertices can be expected to be `independent,' for as we noted in section \ref{programs}, original vertices are not in general to be  thought of as probabilistic. Discussions of fine-tuning in physics often presuppose (either implicitly or explicity) some choice of measure relative to which matches between the values of constants or other original vertices can be understood as `unlikely' or `not typical,' but this approach is controversial \cite{McCoy2017-MCCCTA-5,Hossenfelder_2019} since there is no possibility of empirically verifying such a measure and it is not straightforward to understand what could possibly ground or justify such a thing. So the term `independent' in our definition shouldn't be understood in terms of probabilistic independence - it has a less precise meaning, referring simply to the fact that   by writing some vertices as original vertices, we are stipulating that there is nothing more to be said about their origins - so in particular, there is no connection between them nor any mechanism which adjusts them to match each other. And thus, since there is ex hypothesi no connection between  original vertices, it is reasonable for us to be surprised when we find that that their values are nonetheless very closely matched.  The point of all this  is that when we find this kind of fine-tuning,  we  may be inclined to suspect that there is a deeper explanation for the matching of these vertices - there may in fact be some mechanism which \emph{does} adjust them to match each other, in which case we may wish to expand our generative program to include that mechanism. So fine-tuning of this kind can be regarded as an indication that some part of our model may still benefit from further development:  trying to find explanations for the fact that parameters in our models match in surprising ways has historically been a useful way of making scientific progress, so for example, part of the motivation for Einstein's formulation of GR was the desire to explain the apparent coincidence of the matching values of the gravitational and inertial mass \cite{Lehmkuhl2021-LEHTEP}. Thus looking for `fine-tuning' is potentially a valuable way to decide where to focus our research efforts.

However, we emphasize that in the absence of any objectively correct measure over the values of original vertices, fine-tuning must ultimately be understood as a subjective heuristic rather than an objective flaw of the model. For without a measure there is no   way to quantify how close two original vertices  are or how surprised we should be by their matching - we  cannot make claims to the effect that their matching is `highly improbable,' or `atypical' because those terms assume a well-defined probability distribution over the values of the original vertices. Moreover, even if the vertices are suspiciously close, there is always a possibility that their apparent matching   is really just a coincidence. So  the presence of what appears to be `fine-tuning' doesn't \emph{guarantee} that there will be a deeper explanation to be found - the only way to know for sure is to actually look for an explanation and see if that effort leads to scientific progress.

It is also helpful to consider the relation between fine-tuning in this sense and fine-tuning as it is used in the causal modelling paradigm. Causal graphs which fail to obey `faithfulness' as described in section \ref{general} are sometimes said to be `fine-tuned,' and we can understand the connection between these two notions of fine-tuning by observing that in order to have a causal model where there is a causal arrow from one variable to another   but those variables appear to be statistically independent, it is necessary to `match up' a variety of supposedly independent parameters of the structural equations defining the causal model, in order to ensure that the causal influences exactly cancel out in every possible configuration. So if we write down a generative program in which the parameters of the causal model are  considered as `original vertices,' then fine-tuning of the causal modelling type entails that the generative program will be fine-tuned. Of course, in most real situations (particularly when we are using causal models to describe causal phenomena in the macroscopic world) the parameters of the causal model will not really be original vertices and so there may be opportunities for them to arise through some dynamical or non-dynamical process which explains the connection between them - for example, this occurs in `equilibration' explanations of fine-tuning, as discussed in ref  \cite{articleDD}.  However, when we are dealing with `fine-tuning' not in ordinary causal models but in graphs encoding ontological priority of other kinds, it may be more  plausible that the parameters of the causal model should genuinely be considered as original vertices, so failures of faithfulness may indeed indicate that the generative program is fine-tuned.

\begin{figure}
	\centering
	\begin{tikzpicture}[scale=0.7]

			\coordinate (cp) at (0.2,0);
			\coordinate (cm) at (-0.2,0);

	\coordinate (c) at (0,0); 
	\coordinate (b) at (-1,-1);  
    \coordinate (bb) at (-0.7,-1.3);  
	\coordinate (a) at (1,-1); 
   \coordinate (aa) at (0.8,-1.3);  
	
	\node[above, black] at (c) { $C$}; 
	\node[below, black] at (b) { $C_o$}; 
	\node[below, black] at (a) { $\Lambda$}; 
	
	\draw[black,   arrows={-triangle 90}] (b) -- (cm); 
	\draw[black,   arrows={-triangle 90}] (a) -- (cp); 
 	\draw[black,   arrows={-triangle 90}] (bb) -- (aa);

	\end{tikzpicture}	
	\caption{Alternative structure of ontological priority determining the value of the cosmological constant.}
	\label{cosmo}
\end{figure}
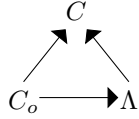

We can also use the relation between these two types of fine-tuning to observe that when we have a fine-tuned generative program, we cannot always overcome that problem by simply switching to a model in which one of the original vertices is no longer original, because that is liable to give rise to a DAG which does not obey faithfulness and thus we will still have a fine-tuning problem.  For example, suppose we try to solve the fine-tuning problem for the cosmological constant  \cite{https://doi.org/10.48550/arxiv.2112.12235} by postulating a   generative program as represented by the DAG shown in figure \ref{cosmo}, where  the bare cosmological constant $C_o$ is an original vertex but the  cut-off scale $\Lambda$  is no longer an original vertex - instead it is a function of $C_o$, calculated in a way which ensures that $C$ will always remain small. Subsequently   $C_o$ and $\Lambda$ are used to calculate $C$ in the usual way. But this DAG fails to obey faithfulness:   although there are arrows of ontological priority from both $C_o$ and $\Lambda$  to $C$,  $C_o$   and $C$ are  statistically independent, and likewise    $\Lambda$ and $C$ are  statistically independent, since regardless of the value of $C_o$ and $\Lambda$, $C$ will always be very close to zero.  So although we no longer have the problem that the values of $C_o$ and $\Lambda$  precisely match up despite the absence of any connection between them, we instead have the problem that the specific parameters used to calculate the relationships between $C_o$ and $\Lambda$  and $C$ have to match up in a highly specific way to make the dependence relations cancel out, so we have not in fact managed to eliminate fine-tuning by this simple stratagem.

 \section{Free Will}

 The generative program also offers a clear way to formalize and visualize various ongoing debates around free will, because concerns   around `determination' have long been central to the accounts that philosophers give of free will or its absence. Perhaps the most common concern raised about free will is that physics tells us the past determines the future, and thus  human actions are fully determined by conditions in the distant past, which suggests that you are not really responsible for your actions at all - you are no more than a vessel through which the distant past exercises its influence on the present. Ref  \cite{VanInwagen1983-VANAEO} puts the matter thus: `\emph{If determinism is true, then our acts are the consequences of the laws of nature and events in the remote past. But it is not up to us what went on before we were born [i.e., we do not have the ability to change the past], and neither is it up to us what the laws of nature are [i.e., we do not have the ability to break the laws of nature]. Therefore, the consequences of these things (including our present acts) are not up to us.}'

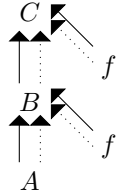
\begin{figure}
	\centering
	\begin{tikzpicture}[scale=0.7]

	\coordinate (c) at (0,0);
	\coordinate (cp) at (0.2,0);
	\coordinate (cm) at (-0.2,0);
	\coordinate (b) at (0,-1);
	\coordinate (bp) at (0.2,-1);
	\coordinate (bm) at (-0.2,-1);
	\coordinate (bp2) at (0.2,-1.6);
	\coordinate (bm2) at (-0.2,-1.6);
	\coordinate (a) at (0,-2.5);
	\coordinate (ap) at (0.2,-2.5);
	\coordinate (am) at (-0.2,-2.5);
	
	\coordinate (i1) at (1.5,-0.3);
	\coordinate (i2) at (1.5,-1.8);
	
	\node[above, black] at (c) { $C$}; 
	\node[below, black] at (b) { $B$}; 
	\node[below, black] at (a) { $A$}; 
	
	\node[below, black] at (i1) { $f$}; 
	\node[below, black] at (i2) { $f$}; 
	
	\draw[black, dotted, arrows={-triangle 90}] (bp) -- (cp);
	\draw[black,   arrows={-triangle 90}] (bm) -- (cm);
	
	\draw[black, dotted, arrows={-triangle 90}] (ap) -- (bp2);
	\draw[black,   arrows={-triangle 90}] (am) -- (bm2);

	\coordinate (c1) at (0.4,0.2);
	\coordinate (c2) at (0.4,0.5);
	
	\coordinate (i11) at (1.2,-0.6);
	\coordinate (i12) at (1.2,-0.3);
	
	\draw[black, dotted, arrows={-triangle 90}] (i11) -- (c1);
	\draw[black,   arrows={-triangle 90}] (i12) -- (c2);
	
	\coordinate (b1) at (0.4,-1.4);
	\coordinate (b2) at (0.4,-1.1);
	
	\coordinate (i21) at (1.2,-2.2);
	\coordinate (i22) at (1.2,-1.9);
	
	\draw[black, dotted, arrows={-triangle 90}] (i21) -- (b1);
	\draw[black,   arrows={-triangle 90}] (i22) -- (b2);

	\end{tikzpicture}	
	\caption{Generative program with original vertices $f$ corresponding to free actions.  }
	\label{fwx}
\end{figure}

\subsection{`Undetermined' free will}

If `determination' is interpreted in causal terms, then one natural way of defining `free will' is to say that an action is free if it is not caused by anything other than the will of the agent, which in turn is `free'  in the sense that it is  not caused by anything at all. However, within the generative program framework we may instead wish to interpret `determination' in terms of ontological priority, which leads to the idea that   an action is `free' if it is, in some appropriate way, associated with an original vertex in a DAG representing the actual ontological priority structure of our world, as depicted in figure \ref{fwx}.   In this section, we will discuss various different ways in which this `undetermined action' conception of free will could be implemented in the generative program framework, focusing on the question of whether there is a consistent way to come up with a DAG modelling a realistic physical theory in which human actions are associated with original vertices.  
 
\begin{figure}[!ht]
		\centering
		\begin{tikzpicture}[scale=0.7]
		
		\coordinate (c) at (0,0);
			\coordinate (cp) at (0.2,0);
			\coordinate (cm) at (-0.2,0);
		\coordinate (b) at (0,-1);
		\coordinate (bp) at (0.2,-1);
		\coordinate (bm) at (-0.2,-1);
			\coordinate (bp2) at (0.2,-1.6);
		\coordinate (bm2) at (-0.2,-1.6);
		\coordinate (a) at (0,-2.5);
		 \coordinate (ap) at (0.2,-2.5);
		 \coordinate (am) at (-0.2,-2.5);
	 
		\node[above, black] at (c) { $C$}; 
			\node[below, black] at (b) { $B$}; 
				\node[below, black] at (a) { $A$}; 
			
				\draw[black, dotted, arrows={-triangle 90}] (bp) -- (cp);
			\draw[black,   arrows={-triangle 90}] (bm) -- (cm);
			
				\draw[black, dotted, arrows={-triangle 90}] (ap) -- (bp2);
			\draw[black,   arrows={-triangle 90}] (am) -- (bm2);
	 
		\end{tikzpicture}	
		\caption{A deterministic theory in which the direction of ontological priority matches the `common sense' direction of causation.}
		\label{dtc}
	\end{figure}
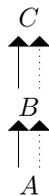

First, if  we are using a graph of ontological priority which coincides with   a standard causal-dynamical picture of reality  where the initial state is ontologically prior to all later events, as those who espouse a direct-access approach to the flow of time will generally be obliged to do, then there are quite strong limitations on the extent to which actions can be `undetermined.'  Indeed, in a deterministic causal-dynamical picture it follows immediately that no actions can be free in this sense, since all of the original vertices are found at the very beginning of time when presumably there are no agents around to take actions, as depicted in figure  \ref{dtc}.

On the other hand, if we now adopt a perspectival approach to the flow of time, in which the experience of conscious agents is understood to supervene only on  the nodes and not on the arrows of the DAG, then our graph of ontological priority is not required to  coincide with a standard causal-dynamical picture of reality, and therefore we have many more options.   For example, consider again a deterministic and time-reversal invariant theory as depicted  in figure \ref{dtc}.   Because the theory is time-reversal invariant, it is equally the case that the \emph{final} state together with the laws of nature can determine the course of history, so if we take it that agents don't have direct access into the process of `coming-into-being' then there is another possible choice of DAG representing ontological priority which gives rise to exactly the same course of history but which has its original vertices at the final condition and all arrows running backwards in time, as depicted schematically in figure \ref{dtb}. So although it is true that according to our usual notion of strong causation, `\emph{our acts are the consequences of the laws of nature and events in the remote past}' it need not be the case that this convention reflects the real structure of ontological priority. 

\begin{figure}
	\centering
	\begin{tikzpicture}[scale=0.7]

	\coordinate (c) at (0,0);
	\coordinate (cp) at (0.2,0);
	\coordinate (cm) at (-0.2,0);
	\coordinate (b) at (0,-1);
	\coordinate (bp) at (0.2,-1);
	\coordinate (bm) at (-0.2,-1);
	\coordinate (bp2) at (0.2,-1.6);
	\coordinate (bm2) at (-0.2,-1.6);
	\coordinate (a) at (0,-2.5);
	\coordinate (ap) at (0.2,-2.5);
	\coordinate (am) at (-0.2,-2.5);
	
	\node[above, black] at (c) { $C$}; 
	\node[below, black] at (b) { $B$}; 
	\node[below, black] at (a) { $A$}; 
	
	\draw[black, dotted, arrows={-triangle 90}] (bp) -- (cp);
	\draw[black,   arrows={triangle 90-}]  (cm) -- (bm);
	
	\draw[black, dotted, arrows={-triangle 90}] (ap) -- (bp2);
	\draw[black,   arrows={-triangle 90}] (bm2) -- (am) ;
	
	\end{tikzpicture}	
	\caption{A deterministic theory in which the original vertices occur on a time-slice which is not at the beginning or end of time - here, B is an original vertex which could correspond to a human action.}
	\label{dtf}
\end{figure}
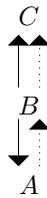

Now, one might object that although there are two valid DAGs representing ontological priority with arrows  going in opposite directions, there is no such DAG in which your action is an original vertex, so either way your action is not an original vertex.  But note that in our deterministic and time-reversal invariant theory, the initial and final conditions are not in any way special: in fact the state \emph{on any time-slice} together with the laws of nature fully defines the course of history, as we can simply evolve both forwards and backwards in time from the given time-slice. So   if we are dealing with a theory of this kind we can always choose a time-slice $T$ containing your action and then formulate a DAG representing ontological priority such that all the original vertices are on the time-slice $T$, with arrows running both forwards and backwards in time from the vertices on $T$, as depicted in figure \ref{dtf}. So, again assuming that agents do not have direct accesss into the process of `coming-into-being,' there is a possible description of the theory in which  your current action is indeed an original vertex, so it is free in the `undetermined' sense.

	\begin{figure}
		\centering
		\begin{tikzpicture}[scale=0.7]

		\coordinate (c) at (0,0);
		\coordinate (cp) at (0.2,0);
		\coordinate (cm) at (-0.2,0);
		\coordinate (b) at (0,-1);
		\coordinate (bp) at (0.2,-1);
		\coordinate (bm) at (-0.2,-1);
		\coordinate (bp2) at (0.2,-1.6);
		\coordinate (bm2) at (-0.2,-1.6);
		\coordinate (a) at (0,-2.5);
		\coordinate (ap) at (0.2,-2.5);
		\coordinate (am) at (-0.2,-2.5);
		
		\node[above, black] at (c) { $C$}; 
		\node[below, black] at (b) { $B$}; 
		\node[below, black] at (a) { $A$}; 
		
		\draw[black, dotted, arrows={-triangle 90}] (bp) -- (cp);
		\draw[black,   arrows={triangle 90-}] (bm) -- (cm);
		
		\draw[black, dotted, arrows={-triangle 90}] (ap) -- (bp2);
		\draw[black,   arrows={triangle 90-}] (am) -- (bm2);
		
		\end{tikzpicture}	
		\caption{A deterministic theory in which the direction of ontological priority is opposite to the `common sense' direction of causation.}
		\label{dtb}
	\end{figure}
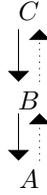

\begin{figure}
		\centering
		\begin{tikzpicture}[scale=0.7]

		\coordinate (past) at (0,0);
		
			\coordinate (consciousness) at (2,2);
			\coordinate (consciousnessplus) at (2.5,2);
			
				\coordinate (action) at (0,4);
				
				\coordinate (rov) at (4,2);
    \coordinate (rov2) at (6,2);

		\node[below, black] at (past) {personal history};  
			\node[right, black] at (consciousness) {decision };  
				\node[above, black] at (action) {action};  
				\node[right, black] at (rov2) { original vertex};  
		
		\draw[black,   arrows={-triangle 90}] (consciousness) -- (action);  
			\draw[black, arrows={-triangle 90}] (past) -- (consciousness);  
				\draw[black,   arrows={-triangle 90}] (rov2) -- (rov);  
		
		\end{tikzpicture}	
		\caption{DAG in which the agent's choice is represented by an original vertex which is subsequently combined with information about physical events.}
		\label{fw4}
	\end{figure}
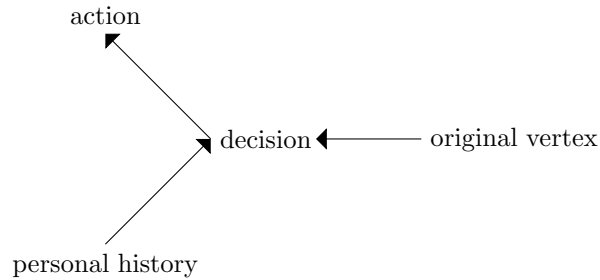

Now, this rewriting only gives you `free will' at a single moment in time. However,  it is possible to imagine   an even more radical rewriting of the DAG where the original vertices are scattered throughout time: this can be done if we can find a set of events across all of history which together contain enough information that the full set of them, along with the time-evolution laws, is enough to determine (but not overdetermine!) the entire course of history. And possibly it could be the case that some such set of events might contain the set of all human actions, or at least a reasonably large subset of human actions. If so, then it may be possible to come up with a DAG in which all of our actions are associated with original vertices, as depicted in figure \ref{fw4}, and thus despite the deterministic and time-reversible evolution, our choices  can truly be regarded as free in the `undetermined' sense. 

One potential danger of scattering `original vertices' across spacetime is that this could potentially give rise to either logical contradictions or fine-tuning. Of course, if one starts from a pre-existing block universe and simply identifies a set of original vertices that can be used to summarise it, the result will never contain any contradictions. But if one imagines a generative program as  starting from a set of independent vertices and then producing the block from them, then in principle contradictions could arise - for example, if two distinct original vertices  both fully determine a third vertex and they assign it different values. And if the values of the original vertices are specifically chosen in such a way as to avoid contradictions, this is liable to look like a kind of fine-tuning, since the vertices will not really be independent any more. So if we are going to scatter original vertices across spacetime, ideally we would like to do this in such a way that the structure of the program guarantees there will not be contradictions. 

One possible strategy is suggested by Loewer's account of free will, which allows that   an action taken in the present can     `influence' both the past and future, where   event E `influences' O if and only if O counterfactually depends on E.  Loewer places strict limitations on the kinds of influences an action can exert on the past - he argues for an approach known as the `Mentaculus,' where we   evaluate counterfactuals by holding constant the \emph{macrostate} at the beginning of time and the probability distribution over microstates compatible with that macrostate, but not the specific initial microstate \cite{doi:10.1142/9789811211720_0001}. This prescription entails that we can in general `influence' microstates in the past, but the macroscopic past history is held constant. Loewer's account is a compatibilist one, but  inspired by his work we can imagine  an `undetermined' version of his view in which human actions are associated with original vertices giving rise to arrows pointing both into the future and to the past. Because of the way in which Loewer limits influences on the past, it follows that the influence of a human action on an event in the past of that action can never be observed by anyone, since human actions can influence only microstates, not macrostates, and if a macroscopic observer were to use a measurement to learn about these influences then the influences would also be changing a past \emph{macrostate}, i.e. the macrostate of the observer's brain. So in this kind of model it is impossible to learn about a past microstaste $M$ in advance of  the action $A$ that influences it, which means we can't perform an observation of $M$ and then use the observation to select $A$. Therefore despite the arrows going in both directions of time, the DAG for the `undetermined' version of Loewer's approach would remain acyclic. Loewer's approach is presumably not the only possible way of achieving this, but it is a useful proof of principle that it is possible to come up with rules which define DAGs with unusual choices of original vertices without running into contradictions.

Of course, the mere existence of such possible DAGs does not lead straightforwardly to the conclusion that human actions are free, because if we adopt the perspectivalist position which suggest that conscious agents have no direct access into the process of coming-into-being, it follows that  we cannot tell empirically which of the various possible ways of assigning ontological priority to events in spacetime is the correct one, and therefore we have no particular reason to think that the correct DAG is one in which our actions are original vertices. But we also have no particular reason to think that the correct DAG is the one where all the original vertices occur at the beginning of time, so for all we know, our actions \emph{could} be the original vertices, in which case they could indeed be free choices. Thus  this approach at least offers the possibility that our actions could be free in the `undetermined' sense, even if it does not guarantee that this is the case. Indeed, since the perspectivalist will argue that changing the structure of the DAG while leaving the empirical data the same cannot change our experience in any way, they might be inclined to suggest that we should regard ourselves as always having free will, because our experience would be exactly the same  if our actions were the original vertices.  

This discussion demonstrates that there is still a possibility of `undetermined' free will even in the context of deterministic laws. But of course our current understanding of quantum mechanics suggests that the laws of our actual world may not be deterministic, in which case we have even more possibilities for scattering original vertices across spacetime - for example, every time there is a random quantum event, we can have an original vertex which determines  the precise outcome of a specific quantum event in order to get a DAG like the one in figure \ref{fwx}. So if it can be shown that random quantum events play an important role in the human brain, it would seem reasonable to assume that human actions always involve at least one original vertex and are thus `free' in the undetermined sense. And in fact there are several ongoing research programs seeking to identify significant quantum effects in the brain - for example Hameroff and Penrose \cite{HAMEROFF1996453} have proposed models for long-term coherence in microtubule structures in the brain, while Kane \cite{Kane+2014+163+182} has argued that butterfly effects may magnify microscopic phenomena to higher scales so they can contribute to decision-making.

However there are some difficulties with simply associating free actions with original vertices  -   it is essentially an expression of event-causal libertarianism and thus is susceptible to the disappearing agent argument \cite{Pereboom2004-PERIOC,Pereboom2012-PERTDA-5,Griffith2010-GRIWAA}. In the terminology of the generative program framework, this argument says that even if there is some original vertex which wholly or partly determines my action, it's unclear that this original vertex has anything much to do with \emph{me}, so  I seem to `disappear' at the moment of actually making the decision.  After all, my character is in large part the result of the past events in the physical world that have shaped my life, together with genetic and biological facts which are likewise a part of the physical world: but any original vertices contributing to my actions are by definition completely independent of all of these things, so those original vertices are surely not   in any meaningful sense a manifestation of me or my will. From this point of view, a `choice' associated with an original vertex is no better than a completely random choice, since it has nothing to do with the actual person involved - they may determine the set of choices along with their relative probabilities, but not the choice itself. James made this point in 1907, arguing that `\emph{If a ‘free’ act be a sheer novelty, that comes not from me, the previous me, but ex nihilo, and simply tacks itself on to me, how can I, the previous I, be responsible? How can I have any permanent character that will stand still long enough for praise or blame to be awarded?}' \cite{Jamespragmatism} Similarly, various philosophers have argued that an action's being caused by appropriate reasons is in fact the essence of free will - for example, Sartorio argues that when we act freely, our conduct is caused  by reasons to act as we do and also the absences of reasons to do otherwise \cite{Sartorio2016-SARCAF-2} - and from this point of view, an original vertex is not actually a meaningful expression of free will since it is by definition not caused by any reasons. 

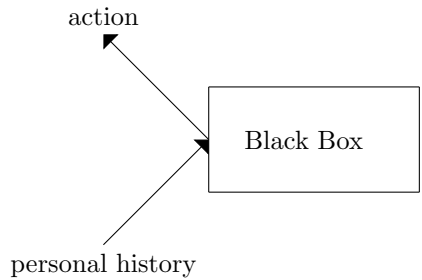
\begin{figure}
		\centering
		\begin{tikzpicture}[scale=0.7]

		\coordinate (past) at (0,0);
		
			\coordinate (consciousness) at (2,2);
				\coordinate (consciousnessplus) at (2.5,2);
				\coordinate (action) at (0,4);
    \coordinate (consciousness2) at (2,2.5);
       \coordinate (consciousness3) at (2,1.5);
	 
		\node[below, black] at (past) {personal history};    
				\node[right, black] at (consciousnessplus) {  Black Box };  
				\node[above, black] at (action) {action};  
\draw[black] (2,1) -- (2,3);
\draw[black] (2,1) -- (6,1);
\draw[black] (2,3) -- (6,3);
\draw[black] (6,1) -- (6,3);
  
		\draw[black,  arrows={-triangle 90}] (consciousness) -- (action);  
			\draw[black,   arrows={-triangle 90}] (past) -- (consciousness);  
		
		\end{tikzpicture}	
		\caption{ DAG in which input from the external world is combined with original information in a vertex representing the free choice of the agent.}
		\label{fw3}
	\end{figure}

For those who are swayed by the disappearing agent argument, it would perhaps seem more natural to   adopt an agent-causal libertarian approach. Here we will attempt to model Clarke's agent-causal approach, which is based on the idea that physical events underlying beliefs and desires are not sufficient in themselves to bring about a free choice but require the agent’s exercise of agent-causal power to make those events causally efficient. This is  represented in figure \ref{fw3}, where we add to the structure of ontological priority a `black box' process, in which some information from previous physical events is synthesized   together with some  free volition in order to arrive at the decision made by the agent. This would still allow the decision to be at least somewhat `free' in the undetermined sense.

Now, it is very natural to ask what exactly is supposed to be going on inside this black box. We can immediately conclude that the black box is not deterministic, because if it were the free volition would not actually be playing any role in deciding the output, and thus this would not represent a meaningful notion of freedom after all. So if we try to expand the contents of the box within the ontological priority framework, it seems we must represent it as something structurally similar to a probabilistic vertex: the inputs to the box are effectively determining a probability distribution, and then the `free volition' is represented by an original vertex which plays the role of actually selecting the output in accordance with the specified probability distribution.  But if that is the case then the DAG of figure \ref{fw3} has the same information flow as the DAG of figure \ref{fw4}, which would entail that agent-causal libertarianism is not really meaningfully different from event-causal libertarianism once we clarify the nature of the information-flow in these models. Thus, if one wishes to maintain that agent-causal libertarianism is meaningfully different from event-causal libertarianism, it seems one must resist the move to expand the black box in this way. So   the agent-causal libertarian must hold that whatever is going on inside the black box, it is neither deterministic nor random, so it apparently cannot be represented at all within the ontological priority framework described here. 

From some points of view this may be regarded as a problem for the agent-causal libertarian: their account depends on postulating a mysterious `black-box' form of information processing, but they cannot tell us anything about how this information processing actually works, and indeed since it is neither deterministic nor random it apparently doesn't correspond to any known type of information-processing.  People who are disinclined to believe in mysterious physical processes which can't be clearly specified may take this as an argument against agent-causal libertarianism, at least until a better account of the operation of the black box can be provided.  On the other hand, agent-causal libertarians might perhaps respond that consciousness is complex and mysterious and perhaps the exact nature of the information processing going on in the black box is just beyond the comprehension of agents like us with our limited cognitive capabilities. We will not attempt to adjudicate between these approaches here, but we think the two  possible representations of agent-causal libertarianism given in figures \ref{fw3} and  \ref{fw4} are helpful to clarify exactly what is at stake in this discussion and why one might wish for further clarification from the agent-causal libertarian.

\subsection{Participatory Free Will}

	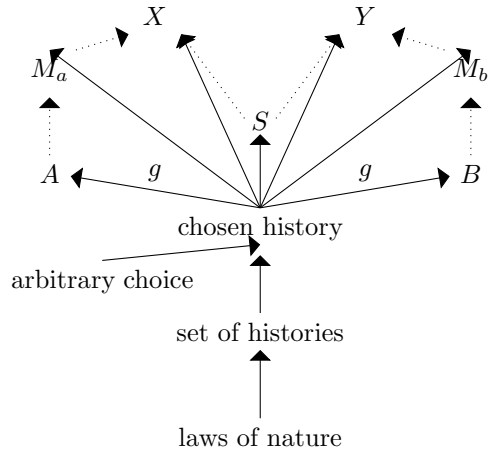
\begin{figure}[!ht]
		\centering
		\begin{tikzpicture}[scale=0.7]

		\coordinate (s) at (0,0);
			\coordinate (sm) at (0.3,-0.3);
				\coordinate (sp) at (-0.3,-0.3);
			\coordinate (smin) at (0,-0.6);
			\node[below, black] at (s) { $S$}; 
		\coordinate (y) at (2,2);
			\node[below, black] at (y) { $Y$}; 
				\coordinate (yp) at (1.5,1.7); 
					\coordinate (yp2) at (1.5,1.3);
						\coordinate (yp3) at (2.5,1.3); 
			\coordinate (ym) at (2.5,1.7); 
		\coordinate (mb) at (4,1); 
			\coordinate (mbm2) at (4,0.1);
				\coordinate (mb3) at (3.5,1); 
			\node[below, black] at (mb) { $M_b$}; 
		\coordinate (b) at (4,-1); 
			\node[below, black] at (b) { $B$}; 
		\coordinate (l) at (0,-2); 
			\coordinate (lp) at (0,-2.7); 
				\coordinate (lpp) at (0,-2.9); 
	\node[below, black] at (l) { chosen history}; 
		\coordinate (l2) at (0,-4); 
			\coordinate (l2p) at (0,-4.7); 
		\node[below, black] at (l2) { set of histories}; 
				\coordinate (l3) at (-3,-3);  
			\node[below, black] at (l3) {arbitrary choice}; 
		\coordinate (l4) at (0,-6); 
		\node[below, black] at (l4) {laws of nature};

				\coordinate (ma) at (-4,1); 
						\coordinate (mam) at (-3.6,-1.4);
		\coordinate (mbm) at (3.6,-1.4);
					\coordinate (mam2) at (-4,0.1);
						\coordinate (ma3) at (-3.5,1); 
		\node[below, black] at (ma) { $M_a$}; 
			\coordinate (x) at (-2,2); 
			\coordinate (xp) at (-1.5,1.7); 
			\coordinate (xp2) at (-1.5,1.3); 
				\coordinate (xm) at (-2.5,1.7); 
				\coordinate (xp3) at (-2.5,1.3); 
		\node[below, black] at (x) { $X$}; 
		\coordinate (a) at (-4,-1); 
		\node[below, black] at (a) { $A$};
		
			\draw[black, arrows={-triangle 90}] (l4) -> (l2p);
			\draw[black, arrows={-triangle 90}] (l2) -> (lpp);
				\draw[black, arrows={-triangle 90}] (l3) -> (lp);
		
					\draw[black, arrows={-triangle 90}] (l) -- (mam);
		\draw[black, arrows={-triangle 90}] (l) -> (smin);
		\draw[black, arrows={-triangle 90}] (l) -- (mbm);
			\draw[black, arrows={-triangle 90}] (l) -> (mb);
	 	\draw[black, arrows={-triangle 90}] (l) -> (ma);
	 		\draw[black, arrows={-triangle 90}] (l) -> (xp2);
	 			\draw[black, arrows={-triangle 90}] (l) -> (yp2);
		
		\draw[black, dotted, arrows={-triangle 90}] (sm) -- (yp2);
		\draw[black, dotted, arrows={-triangle 90}] (sp) -- (xp2);

			\coordinate (a) at (-4,-1); 
					\coordinate (mam2) at (-4,0.1);
		\draw[black, dotted, arrows={-triangle 90}] (a) -- (mam2);
			\coordinate (mam22) at (-4.3,0.1);
				\coordinate (a22) at (-4.3,-1);

			\coordinate (ab) at (4,-1); 
					\coordinate (mam2b) at (4,0.1);
		\draw[black, dotted, arrows={-triangle 90}] (ab) -- (mam2b);
			\coordinate (mam22b) at (4.3,0.1);
				\coordinate (a22b) at (4.3,-1);

			\draw[black, dotted, arrows={-triangle 90}] (ma3) -- (xp3);
				\draw[black, dotted, arrows={-triangle 90}] (mb3) -- (yp3);

					\node[below, black] at (2,-1) { $g$}; 
						\node[below, black] at (-2,-1) { $g$}; 
		
		\end{tikzpicture}	
		\caption{All-at-once model, including  an additional input variable which selects one history from the set of dynamically possible histories, so that history can then be instantiated as the actual course of history.}
		\label{aaom}
	\end{figure}

A completely different approach to free will can be defined within `all-at-once' style models, as advocated in refs \cite{Wharton,Adlamspooky}. Models of this kind  do not postulate temporal evolution or any other directed process by which the course of history is generated: instead the laws specify a set of nomically possible histories and some history from this set is selected and instantiated in an atemporal way. The selection from the set can be thought of as akin to the selection of an initial state in an ordinary time-evolution model: we simply select and actualise the whole history rather than just the initial state. 

In this kind of picture, no event in spacetime is  ontologically prior to any other, as depicted in figure \ref{aaom}, and no set of events is really `the cause' of the rest of history: events at the beginning of time have no better claim to this title than any other set of events. Rather, events happening across all of history stand in mutual, reciprocal relations to one another: there is an underlying structure of ontological priority, which includes some original vertices, but none of the original vertices are located in spacetime. 

So what does this mean for free will? Well, if one is committed to the `undetermined' account of freedom then one will have to conclude that in this picture nobody really has freedom, since no actions are associated with original vertices. But on the other hand, if one's  concerns about free will stem largely from the idea that all our actions are determined by factors outside of our control in the far distant past, then the fact that in these models no direction of time corresponds  to the  objectively correct direction of ontological priority  should assuage those concerns, for this entails that it is no more true to say that our actions are determined by the distant past than it is to say that the past is determined by our actions. This  leads to a conception of freedom which emphasizes the participatory nature of agency: in this picture the course of history is selected in such a way that all moments of time play an equal role, so your current actions are no more or less important than the events at the beginning and end of time. Your choices play a real part in determining the course of history, because all other events must be  adjusted to fit with your choices, just as your choices are mutually adjusted to fit other events. This way of thinking allows for your actions to be `free' in a genuine, ontological sense while also allowing that meaningful choices are not random but rather are appropriately linked with the events in your past that have formed your identity. 

Obviously, this approach in itself is not a definition of free will: the argument applies equally well to all events in history and every action that a person takes, regardless of whether the action corresponds to what we would ordinarily call a `free choice' or whether it is the result of someone holding a gun to the agent's head and making demands. So there is still a need to formulate criteria which can be used to make individual judgements about which actions count as free for moral, legal and religious purposes. However, this approach does serve to defuse the blanket challenge to free will posed by the idea that everything is ultimately determined by the distant past; it allows for the conceptual possibility that at least some human actions may count as `free,' so we can  go on to decide which particular actions we wish to regard as free\footnote{Our own view is that one should not seek a univocal criterion which divides all actions neatly into `free' and `not free' - rather one should acknowledge that human actions are subject to a range of different internal and external constraints, and it is better to focus on understanding the complex interplay of constraints in a specific case  than to assign a simple binary label classifying the action as `free' or `not free'.}.

For example, it is often said that we have free will if it is possible for us to act otherwise, and determinism is supposed to undermine this kind of free will because, if we condition on the microstate of the world up until the time of the action, then only one action is possible. But if no events in spacetime are really ontologically prior to any others, then there is no reason why the whole of the past must be held constant when one assesses what is possible at a given time. Rather we are free to make contextual choices about what factors to hold constant, so for example if we are assessing whether an action is free we will most likely hold constant factors like the presence or absence of external compulsion, intoxicants, phobias etc, but we need not hold constant the exact past microstate. Now, this is a fairly standard compatibilist argument - for example,  List suggests that the relevant notion of possibility in `it was possible to act otherwise' involves conditioning not on the past microstate but on what he calls `the agential state' which describes various psychological factors, and in many cases if we condition only on the agential state then we can conclude it was possible for us to act otherwise, so we can distinguish between cases in which we have free will and cases in which we do not \cite{https://doi.org/10.1111/nous.12019}. The account we have given here adds legitimacy to approaches of this kind, because it entails that these approaches are not simply sweeping under the carpet the fact that the action is really determined by the whole past microstate - in fact all events determine all other events and therefore any meaningful statement of physical possibility must involve a choice about what we are going to hold constant\footnote{This does not apply to statements about physical possibility which refer to the whole universe at once, e.g. saying which entire courses of history are physically possible or impossible. However, evidently such statements are far too general to be applied to questions about free will in any particular case.}. From this point of view, holding only the `agential state,' or some such thing constant  is no more or less correct at a physical level than holding the past microstate fixed, and thus it is perfectly reasonable to say that in some circumstances it is possible for us to act otherwise.

Note that this notion of freedom is similar to the one featuring in  the definition of superdeterminism which we presented in section \ref{ncs}. This definition uses the past empirical experiences of an agent to define the space of possible choices which must be considered as counterfactuals, which is similar to using List's `agential state' to determine which actions were `possible,' or in the terms of   \cite{refining} `empirically plausible': in models which are superdeterministic according to this definition,  the agent's conception of which actions they are free to perform, holding constant only the `agential state,' does not match the reality of what the generative program allows, and hence the program is `conspiratorial'  in the sense that the agent's experiences deceive them into believing that some impossible choices are possible. That said,    even with a superdeterministic generative program, there is still some intermediate sense in which the agent can be free; provided the response function for a given system allows at least two empirically plausible inputs, and specifies empirically plausible outputs for all of them, then the agent is still free to choose among those in this same sense, even if there are other empirically plausible choices that are in fact impossible in the given program.

\begin{figure}
	\centering
	\begin{tikzpicture}[scale=2]

	\coordinate (c) at (0,0);
	\coordinate (cp) at (0.2,0);
	\coordinate (cm) at (-0.2,0);
	\coordinate (b) at (0,-1);
	\coordinate (bp) at (0.2,-1);
	\coordinate (bm) at (-0.2,-1);
	\coordinate (bp2) at (0.2,-1.6);
	\coordinate (bm2) at (-0.2,-1.6);
	\coordinate (a) at (0,-2.5);
	\coordinate (ap) at (0.2,-2.5);
	\coordinate (am) at (-0.2,-2.5);
	
	\coordinate (i1) at (1.5,-0.3);
	\coordinate (i2) at (1.5,-1.8);
	
		\coordinate (f1) at (2.5,0);

	\coordinate (w1) at (-1.5,-0.3);
	\coordinate (w2) at (-1.5,-1.8);
	
	\node[above, black] at (c) { Final}; 
	\node[below, black] at (b) { Projective}; 
 \node[below, black] at (0,-1.25) { outcome}; 
	\node[below, black] at (a) { Initial};

	\node[below, black] at (1.5,-1.9) { Random}; 
	
	\node[below, black] at (w1) { Weak  }; 
\node[below, black] at (-1.5,-0.5)  { reality  };
	\node[below, black] at (w2) { Weak  }; 
 \node[below, black] at (-1.5,-2)  { reality  };
	
	\draw[black, dotted, arrows={-triangle 90}] (bp) -- (cp);
	\draw[black,   arrows={-triangle 90}]  (cm) -- (bm);
	
	\draw[black, dotted, arrows={-triangle 90}] (ap) -- (bp2);
	\draw[black,   arrows={-triangle 90}] (am) -- (bm2);

	\coordinate (c1) at (0.4,0.2);
	\coordinate (c2) at (0.4,0.5);
	
	\coordinate (i11) at (1.2,-0.6);
	\coordinate (i12) at (1.2,-0.3);

	\coordinate (b1) at (0.4,-1.4);
	\coordinate (b2) at (0.4,-1.1);
	
	\coordinate (i21) at (1.2,-2.2);
	\coordinate (i22) at (1.2,-1.9);
	
	\draw[black, dotted, arrows={-triangle 90}] (i21) -- (b1);
	\draw[black,   arrows={-triangle 90}] (i22) -- (b2);

	\coordinate (c1x) at (-0.4,0);
	\coordinate (c2x) at (-0.4,0.2);
	
	\coordinate (i11x) at (-1.2,-0.6);
	\coordinate (i12x) at (-1.2,-0.3);
	
	\draw[black, dotted, arrows={triangle 90-}] (c1x) --  (i11x);
	\draw[black,   arrows={triangle 90-}] (i12x) -- (c2x);
	
	\coordinate (b1y) at (-0.4,-1.4);
	\coordinate (b2y) at (-0.4,-1.1);
	
	\coordinate (i21y) at (-1.2,-2.2);
	\coordinate (i22y) at (-1.2,-1.9);
	
	\draw[black, dotted, arrows={triangle 90-}] (b1y) -- (i21y);
	\draw[black,   arrows={triangle 90-}] (i22y) -- (b2y);
	
	\coordinate (a1y) at (-0.4,-2.5);
	\coordinate (a2y) at (-0.4,-2.8);

	\draw[black, dotted, arrows={-triangle 90}] (a2y) -- (i21y);
	\draw[black,   arrows={-triangle 90}] (a1y) -- (i22y);

	\coordinate (b1y2) at (-0.8,-1.4);
	\coordinate (b2y2) at (-0.8,-1.1);
	
	\draw[black, dotted, arrows={-triangle 90}] (b1y) -- (i11x);
	\draw[black,   arrows={-triangle 90}] (b2y) -- (i12x);

		\coordinate (f21) at (2.1,-1.1);
	\coordinate (f22) at (2.1,-1.4);
	
		\coordinate (b1x) at (0.6,-1.4);
	\coordinate (b2x) at (0.6,-1.1);

	\end{tikzpicture}	
	\caption{Time-symmetric quantum mechanics without free choices.  In this program, both the final and initial boundary conditions are fixed ontologically prior to anything else within space-time, and they define the weak reality throughout space-time.  Then, at events where unitary evolution of the initial state leads to strong interactions and projective collapse, the outcomes are selected at random using the Aharonov-Bergmann-Lebowitz (ABL) rule and the past and future boundary conditions.  The new projective outcome becomes a new boundary condition which redefines the weak reality both before and after, and this process repeats many times, filling in all of the projective outcomes throughout space-time.}
	\label{TSQM}
\end{figure}
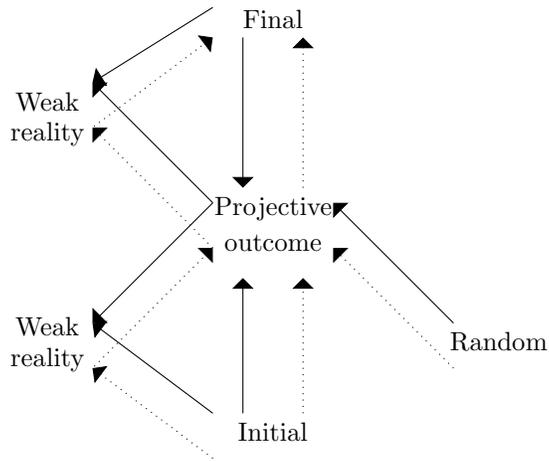

\begin{figure}
	\centering
	\begin{tikzpicture}[scale=2]

	\coordinate (c) at (0,0);
	\coordinate (cp) at (0.2,0);
	\coordinate (cm) at (-0.2,0);
	\coordinate (b) at (0,-1);
	\coordinate (bp) at (0.2,-1);
	\coordinate (bm) at (-0.2,-1);
	\coordinate (bp2) at (0.2,-1.6);
	\coordinate (bm2) at (-0.2,-1.6);
	\coordinate (a) at (0,-2.5);
	\coordinate (ap) at (0.2,-2.5);
	\coordinate (am) at (-0.2,-2.5);
	
	\coordinate (i1) at (1.5,-0.3);
	\coordinate (i2) at (1.5,-1.8);
	
		\coordinate (f1) at (2.5,0);
			\coordinate (f2) at (2.5,-1);

			\node[below, black] at (f2) { Free}; 
	
	\coordinate (w1) at (-1.5,-0.3);
	\coordinate (w2) at (-1.5,-1.8);
	
	\node[above, black] at (c) { Final}; 
	\node[below, black] at (b) { Projective}; 
 \node[below, black] at (0,-1.25) { outcome}; 
	\node[below, black] at (a) { Initial};

	\node[below, black] at (1.5,-1.9) { Random}; 
	
	\node[below, black] at (w1) { Weak  }; 
\node[below, black] at (-1.5,-0.5)  { reality  };
	\node[below, black] at (w2) { Weak  }; 
 \node[below, black] at (-1.5,-2)  { reality  };
	
	\draw[black, dotted, arrows={-triangle 90}] (bp) -- (cp);
	\draw[black,   arrows={-triangle 90}]  (cm) -- (bm);
	
	\draw[black, dotted, arrows={-triangle 90}] (ap) -- (bp2);
	\draw[black,   arrows={-triangle 90}] (am) -- (bm2);

	\coordinate (c1) at (0.4,0.2);
	\coordinate (c2) at (0.4,0.5);
	
	\coordinate (i11) at (1.2,-0.6);
	\coordinate (i12) at (1.2,-0.3);

	\coordinate (b1) at (0.4,-1.4);
	\coordinate (b2) at (0.4,-1.1);
	
	\coordinate (i21) at (1.2,-2.2);
	\coordinate (i22) at (1.2,-1.9);
	
	\draw[black, dotted, arrows={-triangle 90}] (i21) -- (b1);
	\draw[black,   arrows={-triangle 90}] (i22) -- (b2);

	\coordinate (c1x) at (-0.4,0);
	\coordinate (c2x) at (-0.4,0.2);
	
	\coordinate (i11x) at (-1.2,-0.6);
	\coordinate (i12x) at (-1.2,-0.3);
	
	\draw[black, dotted, arrows={triangle 90-}] (c1x) --  (i11x);
	\draw[black,   arrows={triangle 90-}] (i12x) -- (c2x);
	
	\coordinate (b1y) at (-0.4,-1.4);
	\coordinate (b2y) at (-0.4,-1.1);
	
	\coordinate (i21y) at (-1.2,-2.2);
	\coordinate (i22y) at (-1.2,-1.9);
	
	\draw[black, dotted, arrows={triangle 90-}] (b1y) -- (i21y);
	\draw[black,   arrows={triangle 90-}] (i22y) -- (b2y);
	
	\coordinate (a1y) at (-0.4,-2.5);
	\coordinate (a2y) at (-0.4,-2.8);

	\draw[black, dotted, arrows={-triangle 90}] (a2y) -- (i21y);
	\draw[black,   arrows={-triangle 90}] (a1y) -- (i22y);

	\coordinate (b1y2) at (-0.8,-1.4);
	\coordinate (b2y2) at (-0.8,-1.1);
	
	\draw[black, dotted, arrows={-triangle 90}] (b1y) -- (i11x);
	\draw[black,   arrows={-triangle 90}] (b2y) -- (i12x);

		\coordinate (f21) at (2.1,-1.1);
	\coordinate (f22) at (2.1,-1.4);
	
		\coordinate (b1x) at (0.6,-1.4);
	\coordinate (b2x) at (0.6,-1.1);

	\draw[black, dotted, arrows={-triangle 90}] (f22) -- (b1x);
	\draw[black,   arrows={-triangle 90}] (f21) -- (b2x);

	\end{tikzpicture}	
	\caption{Time-symmetric quantum mechanics with agent-causal-libertarian free choice.  This DAG is almost the same as in Fig. \ref{TSQM}, but with the free choices of measurement settings made by agents built into the Hamiltonian and unitary evolution.  The event-causal libertarian case is shown.  For the agent-causal libertarian case, the Free choice would occur within a Black Box with an input arrow the the Initial state, as shown in Fig. \ref{fw3}}
	\label{TSQM_ACL}
\end{figure}
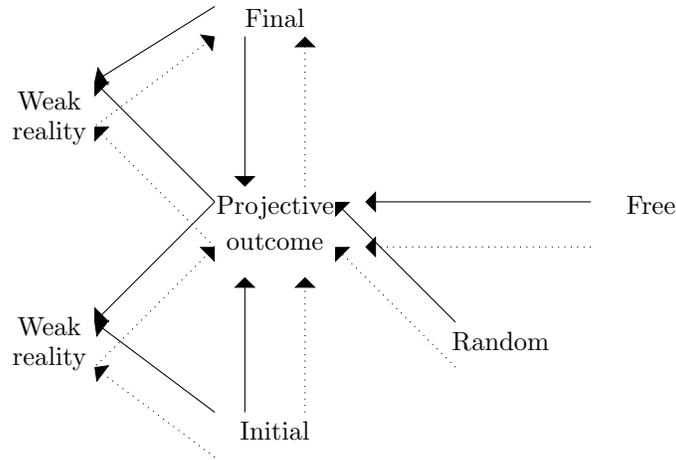

\section{Conclusion}

Within both physics and philosophy there is a long history of linking scientific realism to causation. But this approach is limiting - not every meaningful physical theory can be associated with a set of asymmetric relations linking different physical events, and in particular there is good reason to think that fundamental physics itself contains no asymmetric structures. Thus we would argue that if scientific realism is to continue being relevant to modern physics, it must cease to lean so heavily on causal notions. 

Now, there seems to be a concern amongst some realists that giving up on causation means giving up on realism itself - and more specifically, there is a worry that we cannot talk meaningfully about heuristics like locality, superdeterminism, contextuality and so on without using causal notions. But in this article we have argued that this is not so.  Realism does not require a commitment to causal structure in particular; it simply requires a commitment to the idea that the world has \emph{some} objective structure, which can be encompassed in the claim that there is some fact of the matter about how empirical data is generated. To be precise, we suggest that realism involves a commitment to the idea that there exists some objectively correct generative program generating the empirical data composing our reality, which may or may not include substructures which can sensibly be described as causal. Furthermore, we argue  that any theory which postulates an objectively correct generative program is implicitly a realist theory, despite some claims that theories like this are actually operational.
 
Moreover, the generative program framework can be used to offer precise definitions of terms like locality, superdeterminism, contextuality and so on, without appealing to causal notions - these claims are meaningful only  if associated with some specific generative program. These definitions can also be applied to theories which do propose some causal  and/or retrocausal structure, allowing a precise, unambiguous judgements as to whether they are local, superdeterministic and so on, and also making it possible to compare causal theories to non-causal theories within a single unifying framework. Thus we hope that the generative program framework will offer a way forward for scientific realism which can retain the commitment to objective, explanatory structures even if the world is not fundamentally causal.  

Examples of generative programs for a number of well-known physical theories and interpretations of quantum mechanics are given in algorithmic form in the Supplemental Information (SI), along with some discussion of their properties.  These are our best attempts to characterize how proponents of these interpretations describe them, but they are of course invited to provide their own improved versions of these programs.  That is the purpose of this framework.  To illustrate the utility of programs and their DAGs for particularly counterintuitive theories, Figs. \ref{TSQM} and \ref{TSQM_ACL} show two variant generative programs for the interpretation (see the SI) proposed by Aharonov of the time-symmetric quantum mechanics of Aharonov, Bergmann, and Lebowitz  \cite{aharonov1964time}, one version with agent/event-causal libertarianism, and other without.

Another useful feature of the generative program approach is that putting a theory in the form of a generative program requires us to be precise about the role of agents and their relation to the rest of the physical world. This can help to highlight important structural features of theories that can otherwise be obscured by noncommittal language around agency.  For example, we have noted that operationalism, particularly in the context of a theory like quantum mechanics, tends to lead to a picture of disconnected, isolated agents performing experiments with no relation to other agents or past versions of themselves. This is a significant disadvantage of operationalism which is not very evident from the language used by many operationalists in practice. We can also use the generative program framework to offer new clarity on some traditional definitions about free will - for example, it provides a sharp visual depiction of the disappearing agent argument. It can also be used to express an alternative approach to free will that arises in `all-at-once' approaches to physics: freedom arises from the fact that agency is participatory, since an agent's actions at some time are no more or less important than events occurring at the beginning or end of time. 
 
\printbibliography

\end{document}


\maketitle

\section{Example Programs and DAGs}

We now offer some examples of generative programs for various physical theories, including some interpretations of quantum mechanics. We have made our best effort to honestly characterize what the proponents of these different interpretations would write down as a generative program, but if there are alternatives they would prefer, we would invite them to provide their own versions of the program. In particular, we note that in these programs we have chosen to take a robustly realist approach to lawhood, regarding laws as playing an active role in generating empirical data and this representing them as nodes of our DAGs - however, Humeans and others who prefer a more deflationary approach to lawhood might wish to employ different DAGs in which laws do not play an active role. 

As we have noted, operationalists do not necessarily have to be committed to any particular generative program for these theories, but it is necessary to specify some particular program if we are to answer questions like ``Is this theory  local?" or ``Is this theory superdeterministic?"  The generative program framework makes all of these issues transparent and unambiguous.

We begin with examples of theories in which the ontological priority DAG largely matches the common-sense causal description, and then move on to some cases with more counterintuitive DAGs.  Regardless of the order of ontological priority in the program, the experience of the observer, which is generated as part of the empirical data set, is still the forward-in-time experience of the present.

\subsection{Forward-In-Time Classical Mechanics}

There are a number of different formulations of  classical mechanics (e.g. Newtonian, Hamiltonian and Lagrangian mechanics).  However, here we consider the `common-sense' Newtonian formulation where we select an   an initial state and evolve forward in time, consistent with our everyday forward-in-time notions of causality. Evidently these programs can generate only data consistent with classical physics, but there was a time when this included nearly all empirical data about mechanics. 
\begin{enumerate}
    \item Original vertex (laws): Time-evolution laws consistent with classical equations of motion, in the form of an input-output function which takes a state of an entire closed system as input, and generates the state of that system an infinitesimal time later as output.\label{FTCM:1}
    \item Original vertex (arbitrary selection number): A generic piece of information that can be used to select one element out of any distribution.\label{FTCM:2}
    
    \item Initial state vertex: The arbitrary selection picks out an initial state from the set of allowed inputs to the time-evolution law.  This description refers to vertex \ref{FTCM:1} and vertex \ref{FTCM:2}, so the new vertex has input arrows from those vertices in the DAG.  The initial empirical data is encoded in this state.
    \item Infinitesimally time-evolved selection number vertex:  Uses the selection number for the present (initial) state as an input to the time-evolution function, which generates the selection number for the new state after an infinitesimal time evolution as an output at this new vertex.  This description refers to the present state selection number and the time-evolution law, so the new vertex has input arrows from those two vertices.  This step repeats until the end of the universe, if there is one, generating a new vertex at each iteration.  \label{FTCM:3}
    \item Present state vertex: The new selection number at each time step picks out a present state from the set of allowed inputs to the time-evolution law.  This description refers to vertex \ref{FTCM:1} and the most recent vertex \ref{FTCM:3}, so the new vertex has input arrows from those vertices in the DAG.  The present empirical data is encoded in this state.
\end{enumerate}

The time-evolution law naturally has a more detailed sub-DAG which specifies the interactions between different systems during the time evolution process, and other possible details of how the evolution equations themselves arise, but we omit those details here for simplicity and generality.  Also, note that this program treats the initial and present selection number and state symmetrically, so we can always define a sub-DAG where the present selection from the original DAG become the arbitrary original selection for the sub-DAG.

As we have discussed, there is nothing that prevents us from giving a different program which generates the same data, but perhaps starting from a final state and evolving deterministically backward in time using the same time-symmetric deterministic laws, or starting from an arbitrary time-slice and evolving away from it into the past and future, or even starting from a scattered collection of events throughout time and evolving in both directions to fill in the gaps.  We could also have a modal program which generates the entire history outside of space-time, and then the empirical data are instantiated all-at-once in space-time.  Importantly, note that all of these variant programs exist for any deterministic time-reversible evolution law, as in several of the other cases we will consider.

\subsection{Forward-In-Time Nonrelativistic Quantum Mechanics With Collapse}

The `textbook' interpretation of quantum mechanics uses forward-in-time unitary evolution along with some non-unitary collapse events, both consistent with everyday forward-in-time causality.   

\begin{enumerate}
    \item Original vertex (laws): Deterministic time-evolution laws consistent with nonrelativistic unitary quantum  mechanics, along with nondeterministic time-evolution laws for collapse consistent with the Born rule, all in the form of an input-output function which takes a state for an entire closed system as input, and generates the new state an infinitesimal time later as output.  We will not specify exactly what causes a collapse here, except that we know it must occur by the time a macroscopic observer learns of the outcome.\label{FTNQM:1}
    \item Original vertex (arbitrary selection number): A generic piece of information that can be used to select one element out of any distribution.\label{FTNQM:2}
  \item Initial state vertex: The arbitrary selection picks out an initial condition from the set of allowed inputs to the time-evolution law.  This description refers to vertices \ref{FTNQM:1} and \ref{FTNQM:2}, and so the new vertex has input arrows from those two vertices in the DAG.  The initial empirical data follows from any observables for which this state is an eigenstate.
  \item Infinitesimally time-evolved selection number vertex:  Uses the selection number of the present (initial) state as an input to the time-evolution function, which generates the selection number for the new state after an infinitesimal time evolution as an output at this vertex.  This description refers to the present state and the time-evolution law, so this new vertex has input arrows from those two vertices.  Furthermore, if this time evolution involves a collapse, the new vertex also has an input arrow from another original arbitrary selection number vertex, which is used to sample from the Born rule distribution.  This step repeats until the end of the universe, if there is one, generating a new vertex at each iteration.  \label{FTNQM:3}
  \item Present state vertex: The new selection number at each time step picks out a present state from the set of allowed inputs to the time-evolution law.  This description refers to vertex \ref{FTNQM:1} and vertex \ref{FTNQM:3}, so it has inputs arrows from those vertices.  The present empirical data follows from any observables for which this state is an eigenstate, but there is no other empirical data associated with the unitarily evolving state.
\end{enumerate}

\subsection{Nonrelativistic Bohmian Quantum Mechanics}

The Bohm interpretation of quantum mechanics also uses forward-in-time unitary evolution, along with some hidden variables, and is consistent with everyday forward-in-time causality.  Again this program will only generate data consistent with nonrelativistic quantum mechanics, so it is another program for that model.

 \begin{enumerate}
     \item Original vertex (laws): Deterministic time-evolution laws consistent with nonrelativistic unitary quantum  mechanics in the form of an input-output function which takes a wavefunction for an entire closed system as input, and generates the new wavefunction an infinitesimal time later as output.\label{NBQM:1}
    
     \item Original vertex (laws): Deterministic time-evolution laws consistent with nonrelativistic Bohmian guidance equations for a trajectory in configuration space in the form of an input-output function which takes a present configuration and wavefunction for an entire closed system as input, and generates the new configuration an infinitesimal time later as output.\label{NBQM:2}
   
     \item Original vertex (arbitrary selection number): A generic piece of information that can be used to select one element out of any distribution.  \label{NBQM:3}

   \item Initial wavefunction vertex: The arbitrary selection picks out an initial wavefunction from the set of allowed inputs to the wavefunction evolution law.  This description refers to vertices \ref{NBQM:1} and \ref{NBQM:3}, and so the new vertex has input arrows from those two vertices in the DAG.  \label{NBQM:4}
  
     \item Original vertex (arbitrary selection number): A generic piece of information that can be used to select one element out of any distribution.  \label{NBQM:5} 

     \item Initial configuration selection number vertex:  The arbitrary selection picks out an initial value for the initial configuration selection number, which is a particular point in the configuration space of the entire closed system, where the initial wavefunction is not zero.  This description refers to \ref{NBQM:2}, \ref{NBQM:4} and \ref{NBQM:5}, so this vertex has input arrows from those vertices.  \label{NBQM:6}

     \item Initial configuration vertex: The configuration selected by the number in \ref{NBQM:6}.      Macroscopic empirical data follows from the (hidden) configuration.
    
     \item Infinitesimally time-evolved configuration selection number vertex: Uses the selection number of the present (initial) configuration and the present (initial) wavefunction as inputs to the guidance law, which generates the selection number for the new configuration as an output at this vertex.  This vertex has input arrows from the present configuration selection number , the present wavefunction, and the guidance laws \ref{NBQM:2}. \label{NBQM:7}

     \item Present configuration vertex:  The new selection number at each time step picks out a configuration from the set of allowed configurations.  The vertex has input arrows from \ref{NBQM:2} and \ref{NBQM:7}.  Macroscopic empirical data follows from the (hidden) configuration.
    
   \item Infinitesimal time-evolved wavefunction selection number vertex:  Uses the selection number for the present wavefunction as an input to the time-evolution law, which generates the selection number for new wavefunction after an infinitesimal time evolution as an output at this vertex.  This vertex has inputs from the present wavefunction and the unitary evolution laws, but not the present configuration.  \label{NBQM:8}  Steps \ref{NBQM:7} and \ref{NBQM:8} repeat in sequence until the end of the universe, if there is one, generating new vertices at each iteration. 

   \item Present wavefunction vertex.  The new selection number at each time step picks out a present wavefunction from the set of allowed wavefunctions.  This vertex has inputs from \ref{NBQM:1} and \ref{NBQM:8}.  
 \end{enumerate}

\subsection{Nonrelativistic Everettian Quantum Mechanics}

The Everett interpretation of quantum mechanics uses forward-in-time unitary evolution consistent with everyday forward-in-time causality.  There is no collapse into a unique outcome in this program and instead all possible outcomes are observed by distinct copies of the observer, each occupying one of many worlds.  There are many versions of this interpretation, so we will attempt to be as neutral as possible in how we frame the program.  Again this program will only generate data consistent with nonrelativistic quantum mechanics, so it is yet another program for that model.
\begin{enumerate}
    \item Original vertex (laws): Deterministic time-evolution laws consistent with nonrelativistic unitary quantum  mechanics in the form of an input-output function which takes a wavefunction for an entire closed system as input, and generates the new wavefunction an infinitesimal time later as output.\label{NEQM:1}
    \item Original vertex (arbitrary selection number): A generic piece of information that can be used to select one element out of any distribution.\label{NEQM:2}
  \item Initial wavefunction vertex: Uses the selected number to choose an initial wavefunction in the configuration space of the entire closed system and generates that wavefunction at this vertex.  This description refers to vertices \ref{NEQM:1} and \ref{NEQM:2}, and so the new vertex has input arrows from those two vertices in the DAG.\label{NEQM:3}
  \item Infinitesimal time-evolved wavefunction selection number vertex:  Uses the present (initial) wavefunction selection number as an input to the time-evolution law, which generates the new wavefunction selection number after an infinitesimal time evolution as an output at this vertex.  This vertex has input arrows from the present wavefunction selection number and the unitary evolution laws.  This step repeats until the end of the universe, if there is one, generating new vertices at each iteration.\label{NEQM:4}
 \item Present wavefunction vertex: The new selection number at each time step is used to pick out a wavefunction consistent with the time-evolution laws.  This vertex has input arrows from \ref{NEQM:1} and \ref{NEQM:4}.  Even though the wavefunction never collapses, a macroscopic observer still experiences empirical data consistent with a collapse when they measure a quantum superposition, since after the measurement there is a distinct copy of the observer who sees each possible outcome (and does not see the other copies or outcomes).
\end{enumerate}
The space-time state realism (SSR) model uses virtually the same generative program, with a mathematical formalism in terms of density matrices that is equivalent to the nonseparable universal wavefunction treatment.

\subsection{Forward-In-Time Special-Relativistic Mechanics }

Special relativity is an interesting example because it is usually understood as a `principle theory.' This terminology is due to Einstein, who differentiated between principle theories which use a set of well-confirmed empirical generalizations to make inferences about some phenomena, and constructive theories  which set out a specific model for the reality underlying the phenomena.

In general, a principle theory cannot be used on its own to arrive at some empirical data: it must be supplemented with a constructive theory specifying the nature and behaviour of the systems to which the principles are supposed to apply. This means that  principle theories like special relativity are not really physical theories in the sense we are using the term, and therefore they can't be represented as generative programs.

However there do exist constructive theories that are special relativistic, i.e. they belong to the class of theories satisfying the principles of special relativity - for example, relativistic mechanics and relativistic electrodynamics. We will focus on relativistic mechanics here. This theory starts from  an initial state and evolves forward in time, consistent with everyday forward-in-time causality.  Each observer moves on their own world-line through space-time, and their present observation is of their own past light cone, so two observer at different events can never agree on their empirical descriptions of the world.  This program generates data consistent with special relativity, but not general relativity or quantum mechanics.
\begin{enumerate}
    \item Original vertex: Time-evolution laws consistent with special relativistic mechanics in the form of an input-output function which takes the state on or just within the past light cone of an event as an input, and generates the state at that event as an output.  This type of evolution is required to obey forward-in-time local causality.
    \item Original vertex (arbitrary selection number): A generic piece of information that can be used to select one element out of any distribution.
    \item Initial state vertex: Uses the selection to choose an initial state consistent with the allowed inputs to the time-evolution law.  This means that the initial state can be specified on any complete space-like hypersurface, and is understood to truncate the past light cones used in the time-evolution law.
    \item Infinitesimally time-evolved state selection number vertex: If the state selection number has already been determined at all of the necessary events on or just within the past light cone of an event to form an input to the time-evolution law, then that input is fed into the time-evolution law, and the output state selection number is assigned to a vertex for that event.  This process happens in arbitrary ontological order whenever the input for a given event is fully determined, repeating until states are selected at all events in space-time.  
    \item Event state vertex:  The new selection number at each event picks out the event state from the set of states consistent with the time-evolution laws.  The present experience of an observer is the empirical data from the event states on their past light cone.

\end{enumerate}
The time evolution law that generates the state at a given event must use the states of events within the past light cone to generates the proper time evolution.  To be clear on what Lorentz symmetry says about this, consider that in a particular inertial frame, we would expect to determine the state at that event using the simultaneous states on the space-like surface of the moment just before the event, bounded by the event's past light cone.  If we now boost to other frames, the events on that space-like surface are no longer simultaneous, and there are frames where they are arbitrarily far in the past along those light cones and arbitrarily close to the event.  So, either the states on any such space-like surface are sufficient to compute the state at the new event, or perhaps one needs to use the states on all such surfaces.  This is only in question because the standard methods of defining infinitesimal neighborhoods in calculus do not take special relativity into account, and thus there is nothing built into space-time PDEs that prevents an event from being affected by states (infinitesimally) outside its past light cone, which naturally allows for influences to propagate at any speed.

There are naturally retrocausal versions of this program where events are determined using states in their future light cones, which are bounded by a final state on a space-like hypersurface, and likewise programs which uses a mixture of the causal and retrocausal evolution laws, and all-at-once versions.

\subsection{Local Many Worlds Quantum Mechanics}

The case of quantum mechanics is also interesting - it is not a principle theory, but it is a \emph{framework} theory, i.e., it provides a very general framework within which we can model many different types of systems. As with principle theories, the framework will not predict anything on its own: it must be supplemented with specific details of the kinds of systems to which we are applying quantum mechanics, such as their Hamiltonians and the structure of their state space. Thus when quantum mechanics is represented as a generative program it should in principle include nodes encoding this kind of supplementary information; however, for simplicity we will largely omit these nodes in what follows.

Local many worlds interpretations of quantum mechanics combine the restrictions of special relativity with the wavefunctions of quantum theory.  This model takes an initial state and evolves it forward in time, consistent with everyday forward-in-time causality.  Each observer moves on their own world-line through space-time, and their present observation is of some macroscopic information on their past light cone, so two observers can only agree on their empirical descriptions of their worlds if they are present at the same event.  The observer is constituted from local wavefunctions that also move on world-lines and which evolve by local unitary operations along those world-lines.  A local wavefunction is fully contained within an event, and is only related to events in its past light cone - so there are no spatially extended quantum states in this model.  There is also no collapse into a unique outcome in this program and instead all possible outcomes are observed by distinct copies of the local observer, each occupying one of many local worlds, and the distinct copies' world-lines may then diverge from one another.  This program will generate data consistent with special-relativistic quantum mechanics and quantum field theory.

\begin{enumerate}
    \item Original vertex: Time-evolution laws consistent with local special-relativistic quantum mechanics in the form of an input-output function which takes the state on or just within the past light cone of an event as an input, and generates the state at that event as an output.  This type of evolution is required to obey forward-in-time local causality.  If multiple world-lines converge at an event, this law generates the evolution of all of them.
    \item Original vertex (arbitrary selection number): A generic piece of information that can be used to select one element out of any distribution.
    \item Initial state vertex: Uses the selection to choose an initial state consistent with the allowed inputs to the time-evolution law.  This means that the initial state can be specified on any complete space-like hypersurface in the form of a distinct local wavefunction at each event, and this surface is understood to truncate the past light cones used in the time-evolution law.
    \item Infinitesimally time-evolved state selection number vertex: If the local wavefunctions have already been selected at all of the necessary events on or just within the past light cone of an event to form an input to the time-evolution law, then that input is fed into the time-evolution law, and the outputted local wavefunction selection number is assigned to a vertex for that event.  This process happens in arbitrary order whenever the input for a given event is fully determined, repeating until states are determined at all events in space-time.  Furthermore, after a quantum superpositions is measured by a macroscopic observer at an event, there is a distinct copy of the the observer for each possible outcome, and each copy sees just their one outcome as empirical data - with all such copies are represented within the local wavefunction at that event.
    \item Event state vertex:  The new selection number at each event picks out the event state from the set of states consistent with the time-evolution laws.  Even though the local wavefunctions never collapse, a macroscopic local observer still experiences empirical data consistent with a collapse, since after a measurement there is a distinct copy of the observer who sees each possible outcome (and does not see the other copies outcomes).  The complete present experience of an observer (copy) is the empirical data from the event states on their past light cone, consistent with a particular collapsed outcome for each macroscopic system.
\end{enumerate}

It should be stressed that because of the local wavefunctions evolving on their own world-lines, this is inherently a local hidden variable theory.  However, because this is also a many-worlds theory, the fact that a local hidden variables model generates empirical data consistent with quantum entanglement correlations does not conflict with Bell's theorem (ADD REFS).

\subsection{Aharonov's Time-Symmetric Free Will}

This model is not actually published anywhere else, and represents our best attempt to characterize the order of ontological priority at play in Yakir Aharonov's time-symmetric view of free will, based on numerous discussion with him about this subject.

This model allows for many variations, but the version we present here is consistent with choices being made in real time by an observer who experiences time in the usual forward direction.  In this model, both the initial state and the final state of the universe are fixed ontologically prior to any other intermediate states (and they are not orthogonal).  The standard Hamiltonian is modified to include free inputs from 
agents, so macroscopic things, like measurement settings, can be freely selected, consistent with agent-causal libertarianism.  When a projective measurement occurs, the outcome is determined randomly using the ABL (Aharonov-Bergmann-Lebowitz) rule instead of the Born rule, which uses the final state and the most recent projective outcome state as inputs (with unitary evolution applied to find their values at the present moment), and always determines a result compatible with the final state.  In this way, observers are empowered with truly free choices that affect the way the present plays out, even though the final state of the universe remains fixed.  The fixed initial and final states do not pick out a fixed history for the entire universe, and the observers' freedom allows them to exert some control on the particular history that occurs on the way to the final state.  Note that an observer who performs an ensemble of measurements on identically prepared systems will still find Born rule statistics for the ensemble, so this picture is empirically consistent.

This is ultimately an agent-causal libertarian conception of free will, but the connection to time is extremely counterintuitive.  In this picture, the past and future boundary conditions are both fixed ``first'' in the ontological sense, and the agent's choice at each moment is whether to perform a measurement at all, and if they do, whether it will be a weak or projective measurement, and what the measurement setting will be.  Only if they choose to perform a projective measurement will a collapse occur (using ABL probabilities), and then the outcome becomes the new past boundary condition for future choices (which defines the weak values, and the ABL rule).

\begin{enumerate}
    \item Original vertex (laws): Time-evolution laws consistent with nonrelativistic unitary quantum mechanics and nondeterministic collapse, in the form of an input-output function which takes a wavefunction for an entire closed system as input, and generates the new wavefunction an infinitesimal time later or earlier as output.  Agents' choices provide another nondeterministic part of this evolution law, which affect macroscopic degrees of freedom, but not collapse outcomes.  These agent choices are of the black-box type that are not generally considered original vertices.\label{YAFW:1}
    \item Original vertex (arbitrary selection number): A generic piece of information that can be used to select one element out of any distribution.\label{YAFW:2}
  \item Initial wavefunction vertex: Uses the selected number to choose an initial wavefunction in the configuration space of the entire closed system and generates that wavefunction at this vertex.  This description refers to vertices \ref{YAFW:1} and \ref{YAFW:2}, and so the new vertex has input arrows from those two vertices in the DAG.\label{YAFW:3}
  \item Original vertex (arbitrary selection number): A generic piece of information that can be used to select one element out of any distribution.\label{YAFW:6}
  \item Final wavefunction vertex: Uses the selected number to choose a final wavefunction in the configuration space of the entire closed system and generates that wavefunction at this vertex.  This description refers to vertices \ref{YAFW:1} and \ref{YAFW:6}, and so the new vertex has input arrows from those two vertices in the DAG.\label{YAFW:5}.
  \item Infinitesimal time-evolved wavefunction selection number vertices:  Uses a wavefunction selection number as an input to the time-evolution law, which generates a new wavefunction selection numbers following an infinitesimal time evolution (forward or backward) as an output at this vertex.  If this evolution involves a collapse, the new vertex also has an input arrow from another original arbitrary selection number vertex, which is used to sample from the ABL rule distribution given by the forward-evolved initial state and backward-evolved final state.  These vertices generally have input arrows from both past and future wavefunction selection numbers and the evolution laws.  This step repeats throughout the universe, generating new vertices at each iteration so that all of time is filled in with selections in both directions.\label{YAFW:4}
 \item Present wavefunction vertices: The new selection numbers at each time step are used to pick out wavefunctions consistent with the time-evolution laws.  These vertices have input arrows from \ref{YAFW:1} and \ref{YAFW:4}.
\end{enumerate}

\subsection{Modal All-At-Once Classical Mechanics}

The Classical Mechanics example, and the variants where the order of everyday forward-in-time causality does not match the order of ontological priority, still involved some type of ordered deterministic time-reversible evolution between events, starting with some type of boundary condition in space-time.  Modal programs generate the set of entire histories of the closed system using the same time-reversible deterministic laws, and then one is selected prior to the empirical data being generated in space-time.

\begin{enumerate}
    \item Original vertex: Time-evolution laws consistent with classical equations of motion, in the form of an input-output function which takes any informationally complete boundary condition for an entire closed system as input, and generates the history of the system throughout all of space and time as output.
    \item Set of histories vertex: The time-evolution law is used to generate the entire history corresponding to every allowed input in its domain.\label{MACM:1}
     \item Original vertex (arbitrary selection number): A generic piece of information that can be used to select one element out of any distribution.\label{MACM:2}
    \item Chosen history vertex: Uses the selected number to choose an entire history from among the set, which fixes the events throughout that history all at once.  This description refers to vertices \ref{MACM:1} and \ref{MACM:2}, and so the new vertex has input arrows from those two vertices in the DAG.
    \item Empirical data vertices: The empirical data corresponding to the chosen history is generated throughout space-time all at once.  These vertices all have an input arrow from the chosen history vertex, but there are no arrows between events in space-time.
\end{enumerate}

\subsection{Nonrelativistic All-At-Once Bohmian mechanics}

These are the formal programs for the two different variants of the all-at-once models of Bohmian mechanics - one local but not superdeterministic, and the other superdeterministic but not local.

\begin{enumerate}
    \item Original vertex: Time-evolution laws consistent with nonrelativistic quantum unitary dynamics, in the form of an input-output function which takes a complete wavefunction at some time $t$ (this is a delocalized object defined across the entire space-like hypersurface through $t$, in a particular inertial frame), and generates the history of the wavefunction throughout all of space and time as output.
    
    \item Set of wavefunction histories vertex: The time-evolution law is used to generate the wavefunction history corresponding to every allowed input in its domain.\label{MABM:1}
    
    \item Original vertex: A fixed or random selection of a number in the range 0 to 1.\label{MABM:2}
    
    \item Chosen wavefunction history vertex: Uses the selected number to choose an entire wavefunction history from among the set, which fixes the events throughout that history all at once.  This description refers to vertices \ref{MABM:1} and \ref{MABM:2}, and so the new vertex has input arrows from those two vertices in the DAG.\label{MABM:3}
    
    \item Original vertex: Deterministic time-evolution laws consistent with nonrelativistic Bohmian guidance equations for a trajectory in configuration space, in the form of an input-output function which takes a configuration and a wavefunction at time $t$ as input, and generates the history of that configuration throughout space and time as output.\label{MABM:4}
    
    \item Set of configurations vertex: The guidance law along with the chosen wavefunction history is used to generate the set of all configuration histories using every configuration where the chosen wavefunction history is nonzero.  The vertex had input arrows from \ref{MABM:3} and \ref{MABM:4}. \label{MABM:5}
    
    \item Original vertex (arbitrary selection number): A generic piece of information that can be used to select one element out of any distribution. \label{MABM:6}
    
   \item Chosen configuration history: Uses the selected number to choose one configuration history from the set.  This vertex has inputs from \ref{MABM:5} and \ref{MABM:6}.
   
    \item (nonlocal nonsuperdeterministic variant) Empirical data vertices: The empirical data corresponding to the chosen configuration history is generated throughout space-time all at once.  All of these vertices have an input arrow from the chosen configuration history vertex, and there are no arrows between events in space-time.
    \item (local superdeterministic variant) Empirical data vertices: The entire trajectory of each particle is taken from the chosen history and assigned as a separate local guidance equation to that particle alone.  Each particle moves through space-time on its own scripted trajectory, generating empirical data as it goes, and there are no interactions between different particles.  Vertices are generated at events along that trajectory as the particle moves through space-time such that each event has an input arrow from the one before it, and the first event in that trajectory has an input arrow from the chosen history vertex.
\end{enumerate}
A slight modification of this treatment would generate the set of all wavefunction histories, and then the set of all configuration histories for each wavefunction history, and only then begin to make a random selection.

\subsection{The Parisian Zig-Zag}
We can also consider programs that only generate the data for a particular experimental scenario. The Parisian Zig-Zag is a retrocausal program for a Bell experiment, in which future action results in changes to the past microstate, just as in Loewer's Mentaculus.  For a Bell experiment where one particle from an entangled pair is sent to Alice, and the other is sent to Bob, and they perform freely or randomly selected measurements of their particles at space-like separation, the Zig-Zag DAG is asymmetric, and we arbitrarily choose an ontological priority DAG where the arrow from the vertex where the entangled pair was created leads to Alice and not Bob.
\begin{enumerate}
    \item Original vertex: Deterministic time-evolution laws consistent with nonrelativistic unitary quantum  mechanics, along with nondeterministic time-evolution laws for collapse consistent with the Born rule, all in the form of an input-output function which takes a state for an entire closed system as input, and generates the new state an infinitesimally short time in the past or future as output.  We will not specify exactly what causes a collapse here, except that we know it must occur by the time a macroscopic observer learns of the outcome.
    \item Bell pair preparation vertex: The reduced density matrix for one particle is created at the source and becomes the state of the particle that is sent to Alice.
    \item Alice's measurement selection vertex: A vertex which may or may not be original which specifies a choice of measurement setting
    \item Original vertex (arbitrary selection number): A generic piece of information that can be used to select one element out of any distribution.
    \item (Alice's measurement outcome): Alice's measurement setting, the reduced density matrix from her particle, and the selected number, are fed into the input-output function, which generates a measurement outcome and a corresponding collapsed state for her particle, where the selection number is used to sample from the Born rule distribution for the state and setting.
    \item (Retrocausal Bell pair vertex):  Alice's outcome state is sent retrocausally to the source event, and then the singlet state is projected onto her outcome, which becomes the state of the particle sent to Bob.  This vertex is colocated at the source event, but it is separate from the Bell preparation vertex.  This vertex has inputs from the preparation of the singlet, and from Alice's outcome.
    \item (Bob's measurement selection vertex): A vertex which may or may not be original which specifies a choice of measurement setting
    \item Original vertex (arbitrary selection number): A generic piece of information that can be used to select one element out of any distribution.
    \item (Bob's measurement outcome): Bob's measurement setting, the state of his particle, and the selected number, are fed into the input-output function, which generates a measurement outcome and a corresponding collapsed state for his particle, where the selection number is used to sample from the Born rule distribution for the state and setting.  This procedure guarantees that the correlations of the singlet state will be obeyed by their measurement outcomes.
\end{enumerate}

\section{Locality and Superdeterminism}

There are several key assumptions in Bell's theorem and the Kochen-Specker theorem which are implicitly assumptions about the ontologically real generative program, which is also to say that they are assumptions about the actual workings of nature.  At the risk of being repetitive, this means that the descriptions in the program represent ontologically real physical objects and mechanisms, while other regularities or relations which are not part of a the given generative program are incidental and irrelevant with respect to these assumptions.

We now consider the assumptions of Locality and No Superdeterminism, and evaluate each of our example programs to check if either or both of these are obeyed.

\begin{enumerate}

\item Forward-in-time classical mechanics:  This program has no rules that prevent faster-than-light motion, which can be understood as a type of influence between space-like separated events, so locality is clearly violated.  Next, since there is no entanglement, the program provides separate descriptions of the present states of all objects in space.  This means that an observer can be separate from a system.  Now, the response function when an observer intervenes upon (interacts with) another is the classical time-evolution law which determines the response of both the system and the observer to the interaction.  This law can take any present configuration of the system and observer as an input, and determines a definite response for any such input, so it does not entail anything about the observer's intervention, and thus this program is not superdeterministic.

\item Forward-in-time nonrelativistic quantum mechanics with collapse:  This program has no rules that prevent faster-than-light motion, and so it violates locality.  Next, the present state in this program can be an entangled state of many systems in many locations, but (most) macroscopic observers in entangled superpositions collapse immediately, in which case their descriptions can be separated from the descriptions of any systems they may intervene upon.  When an observer intervenes upon a system, the response is determined at that moment using the time-evolution law.  This law accepts any present state of the observer and system as input and determines a definite response for any such input, and thus this program is not superdeterministic.

 \item Nonrelativistic Bohmian quantum mechanics: This program uses guidance equations that involve influences between space-like separated events, and so it violates locality.  Next, there is no collapse in this program, so the present wavefunction of a system and observer will generally be entangled and nonseparable, which presents a problem for ruling out superdeterminism.  However, if we condition the wavefunction on the hidden configuration of a macroscopic observer, then we seem to end up with separate descriptions for the observer and system.  When the observer intervenes upon the system, the response is determined at that moment using the time-evolution laws.  These laws accept any present wavefunction and configuration of the observer and system as inputs and determine definite responses for any such input, in which case this program is not superdeterministic.  The superdeterminism question is a bit less cut and dried here, since it requires us to consider the conditioned wavefunction in order to obtain a separable description, but this does seem to keep with the correct intuition.

\item Nonrelativistic Everettian quantum mechanics: The universal quantum state in this program is generally entangled and describes the space-like separated present states of many systems at once, which is a nonseparable description of space-like separated events, and thus violates locality.  Next, there is no collapse in this program, so the present state of a system and observer will generally be entangled and nonseparable, which presents a problem for ruling out superdeterminism.  However, if, for each copy of a macroscopic observer, we condition the universal state on the outcomes they have observed, then relative to each copy of the observer, we can end up with separate descriptions of the observer and system.  When the observer intervenes upon the system, the response is determined at that moment using the time-evolution laws.  These laws accept any present state of the observer and system as inputs and determine definite responses for any such input, in which case this program is not superdeterministic.  Again, entanglement makes the question of superdeterminism subtle, but this seems to be the correct intuition.

Now, leaving aside objects with velocities faster than light, the conclusion that the Everettian program is nonlocal due to entanglement is rejected by some, mainly because they employ different definitions of local than the one we have given here.  As such, we will give some arguments here that it is impossible to come up with a consistent definition of locality that is not violated by any program with nonseparable descriptions of space-like separated events.  One virtue of the generative programs framework is that an ontologically real program clearly identifies all of the physical mechanisms in the theory, which is to say that everything the program describes is physically real.  Likewise, any other program which could generate the empirical data is physically irrelevant when evaluating whether the ontologically real program is local or superdeterministic.

In the case of the Everettian (or space-time-state realism) program, the entangled state is a physical object which is also a nonseparable description of many systems spread across the present space-like hypersurface, so this physical object is a delocalized by its own definition.  Any local intervention on one of the entangled systems results in a change to the entire delocalized state, which constitutes a faster-than-light influence between the intervention and all other entangled systems.  The underlying point is that any objects that are used by the program as part of the process of generating the empirical data must be considered as real physical objects.  rather than passive or extrinsic relations, as in the Socrates-Xanthippe argument, which we now discuss.

Some Everettians have argued that the Everett approach is local because the relations it postulates between spacelike separated objects are only passive or extrinsic relations. As an example, consider the case of widowhood: the moment Socrates dies of drinking hemlock, his wife Xanthippe instantaneously becomes a widow, regardless of how far she is from the event of his death. but the feature of widowhood is an extrinsic relation, so nothing about Xanthippe has physically changed and local causality has not been violated.  One might argue that entanglement relations in Everett are extrinsic relations of the same type as Xanthippe's widowhood, since they only need to be used as inputs by the time-evolution law at the time when the two entangled systems meet at an event, and at that point they are not acting non-locally. However, using the generative program framework we can show that there is a disanalogy here.   The standard classical generative program that we would use to describe the macroscopic doings of Socrates and Xanthippe makes no use of the relation `widowhood' to generate any empirical data, which means that in this program, this relation does not need to be given ontological significance, so  locality is not violated.   On the other hand, in the Everett program, the nonseparable entangled state is part of the  real mechanism used to generate the empirical data, and the dynamics of this physically real object violate local causality.

The Everettian might respond by arguing that  the nonseparable entanglement relations are only needed as inputs to the time-evolution functions when the entangled systems meet at some point in the future, so they are only physical objects at those points, and thus changing the entanglement relations while the systems are spatially separated does not count as a nonlocal change to a physical object.  To see the flaw in this argument, one need only consider how the entanglement relation, which exists continuously in this program, is supposed to arrive at the meeting event where it must become a physical object.  If the program obeys local causality, there must be local hidden variables (even if there are many worlds), so this relation must comprise real physical objects that are carried from the event where they are created to meeting events where they are necessary inputs.  Then, because it is a nonseparable relation of all of the entangled systems, it must be carried by all of them, and thus it is a spatially delocalized physical object when those systems are spatially separated, which violates local causality as above.  The underlying point is that the mathematical description of these relations exist at all times, and there is no grounds to deny that they are also physical objects at all times.

As we will see, the local many worlds program uses separable local hidden variables which perform the same role as the nonseparable relations in Everett, but with fully local dynamics in space-time.

 \item Forward-in-time special-relativistic mechanics:  There are no entangled states or faster-than-light motions in this model, so locality is clearly obeyed - as we would expect given that locality is really just a consequence of special relativity.  Next, the present states of an observer and system are always separate, and when an observer intervenes upon a system, the response is determined at that moment using the time-evolution law.  This law accepts any present state of the observer and system as input and determines a definite response for any such input, and thus this program is not superdeterministic.  This is the canonical program that roots our standard intuitions about locality and free choice.

\item Local many worlds quantum mechanics:  Entangled states in this model are confined within individual events in this program, so there are no delocalized objects in space-time and there is no faster-than-light motion, so locality is obeyed.  Next, the present local wavefunctions of an observer and a spatially separated system are always separate, and when an observer intervenes upon a system at an event, the response is determined at that moment using the time-evolution law.  This law accepts any present local wavefunctions of the observer and system as inputs and determines a definite response for any such input, and thus this program is not superdeterministic.  This program follows the intuition of forward-in-time special relativity most closely, with many local worlds in a single objective space-time.

\item Aharonov's Time-Symmetric Free Will:  This model treats entanglement with the same nonseparable states as textbook quantum theory, so it is manifestly nonlocal.  The model is consistent with agent-causal libertarianism, and its machinery can accept any choice the agent makes as input, so it is not superdeterministic.

 \item Modal all-at-once classical mechanics:  

This program also allows faster-than-light motion, and violates locality.  A constrained version of this program which forbids faster-than-light motion would be local.  Next, the only response function available in this theory is the one that selects between entire histories, so given that the set of histories contains every empirically plausible intervention and response, the theory is not superdeterministic.  Note that this theory can only generate data consistent with classical mechanics, so it cannot reproduce entanglement correlations.

 \item Nonrelativistic modal Bohmian quantum mechanics:  This program allows for faster-than-light motion, and violates locality.  If we consider a version where faster-than-light motion is forbidden but there are still nonlocal guidance equations, then we can find both a local superdeterministic variant and a nonlocal nonsuperdeterministic variant.

In the nonlocal nonsuperdeterministic variant, the response functions are entire histories, and these contain every empirically plausible intervention and response, so the theory is not superdeterministic.  However, each history is computed using nonlocal guidance equations and nonseparable wavefunctions, so the theory is manifestly nonlocal.

For the local superdeterministic variant, after choosing a hisotry, the description of each system's fixed trajectory along its world-line is assigned to it as a local hidden variable in space-time, so there are no non-separable descriptions of space-like separated events and theory is local.  But now each trajectory is fixed by its own local hidden variable and there is no interaction between different systems in space-time, so each systems response function effectively has just one intervention and response, and thus the theory is manifestly superdeterministic.  To be explicit, once the history is computed using the nonlocal guidance equations and then trajectories are extracted, we can no longer change the trajectories individually without violating the equations.

 \item Parisian Zig-Zag:  This program creates single-particle states which are then carried locally by the individual particles forward and backward in time along their world-lines, so the states are local hidden variables, and the theory is local. Next, the observers and systems in this model have separate descriptions, and when Alice or Bob measure their particle, the response is determined at that moment using the time-evolution law.  This law accepts any measurement setting and particle state as inputs and determines a definite response for any such input, and thus this program is not superdeterministic.  This theory violates the \textit{No Retrocausality} assumption which forbids ontological priority DAGs with any arrows, or string of arrows, that lead from one event in space-time to another event in its past.

\end{enumerate}